\documentclass{ieeeaccess}

\usepackage{cite}
\usepackage{amsmath,amssymb,amsfonts}
\usepackage{algorithmic}
\usepackage{textcomp}

\usepackage[margin=1in]{geometry}
\usepackage{times,latexsym}
\usepackage{url}
\usepackage[T1]{fontenc}

\usepackage{array}  
\usepackage{tabularx} 
\usepackage{multirow}  
\usepackage{xspace}
\usepackage{booktabs}
\usepackage{graphicx}
\usepackage{xcolor}
\usepackage[hidelinks]{hyperref}

\newcommand{\todo}[1]{}
\newcommand{\bonus}[1]{}
\newcommand{\frejm}[1]{}
\newcommand{\note}[1]{}

\newcommand{\citinl}[1]{\cite{#1}}
\newcommand{\citnoun}[1]{\cite{#1}}

\newcommand{\hla}[1]{#1}
\newcommand{\hl}[1]{#1}

\newcommand{\fone}{$F_{1}$\xspace}

\newcommand{\ldamodel}{LDA\xspace}
\newcommand{\aldamodel}{aLDA\xspace}
\newcommand{\pypmodel}{PYP\xspace}
\newcommand{\nmfmodel}{NMF\xspace}

\newcommand{\rbfgamma}{$\gamma$\xspace}
\newcommand{\autogamma}{$1/NumFeatures$}

\newcommand{\sups}{SupCov\xspace} 

\newcommand{\loned}{L\textsubscript{1}\xspace}
\newcommand{\ltwod}{L\textsubscript{2}\xspace}

\newcommand{\cdcurve}[1]{CD-curve#1\xspace}
\newcommand{\cdc}{AuCDC\xspace}
\newcommand{\cdcm}{AuCDC}

\newcommand{\cosd}{cosd\xspace}

\newcommand{\bst}[1]{\textbf{#1}}
\newcommand{\good}[1]{\underline{#1}}

\newcommand{\bipstab}{InstanceStabil\xspace}
\newcommand{\css}{RefsetStabil\xspace}
\newcommand{\cdcs}{AuCDC-stabil\xspace}

\newcommand{\npmi}{NPMI\xspace}
\newcommand{\cv}{CV\xspace}
\newcommand{\cp}{CP\xspace}
\newcommand{\npmil}[1]{NPMI-#1\xspace}
\newcommand{\cvl}[1]{CV-#1\xspace}
\newcommand{\cpl}[1]{CP-#1\xspace}

\newcommand{\sgmnt}[2]{[#1, #2]}

\begin{document}

\history{Date of publication xxxx 00, 0000, date of current version xxxx 00, 0000.}
\doi{10.1109/ACCESS.2021.3109425} 

\title{A Topic Coverage Approach to Evaluation of Topic Models}

\author{\uppercase{Damir Koren\v{c}i\'{c}\authorrefmark{1},
Strahil Ristov\authorrefmark{1}, Jelena Repar\authorrefmark{2}, and Jan \v{S}najder\authorrefmark{3}}}
\address[1]{Rudjer Bo\v{s}kovi\'{c} Institute, Division of Electronics, 10000 Zagreb, Croatia 
(e-mail: dkorenc@irb.hr, ristov@irb.hr)}
\address[2]{Rudjer Bo\v{s}kovi\'{c} Institute, Division of Molecular Biology, 10000 Zagreb, Croatia
(e-mail: jelena.repar@irb.hr)}
\address[3]{University of Zagreb, Faculty of Electrical Engineering and Computing, 10000 Zagreb, Croatia
(e-mail: jan.snajder@fer.hr)}
\tfootnote{This work was supported in part by the European 
Regional Development Fund under the grant KK.01.1.1.01.0009 (DATACROSS).}

\markboth
{Koren\v{c}i\'{c} \headeretal: A Topic Coverage Approach to Evaluation of Topic Models}
{Koren\v{c}i\'{c} \headeretal: A Topic Coverage Approach to Evaluation of Topic Models}

\corresp{Corresponding author: Damir Koren\v{c}i\'{c} (e-mail: dkorenc@irb.hr).}

\begin{abstract}

Topic models are widely used unsupervised \hl{models capable} of learning topics
-- weighted lists of words and documents -- from large collections of text documents. 
When topic models are used for discovery of topics in text collections, 
a question that arises naturally is how well the model-induced topics correspond to topics of interest to the analyst.
In this paper we revisit and extend a so far neglected approach to topic model evaluation based on measuring topic coverage -- 
computationally matching model topics with a set of reference topics that models are expected to uncover.
The approach is well suited for analyzing models' performance in topic discovery
and for large-scale analysis of both topic models and measures of model quality.
We propose new measures of coverage and evaluate, in a series of experiments, 
different types of topic models on two distinct text domains for which interest for topic discovery exists. 
The experiments include evaluation of model quality, analysis of coverage of distinct topic categories, 
and the analysis of the relationship between coverage and other methods of topic model evaluation.
\hl{
The paper contributes a new supervised measure of coverage, and the first unsupervised measure of coverage.
The supervised measure achieves topic matching accuracy close to human agreement.
The unsupervised measure correlates highly with the supervised one (Spearman's $\rho \geq 0.95$). 
Other contributions include insights into both topic models and different 
methods of model evaluation, and the datasets and code for facilitating future research on topic coverage.
}

\end{abstract}

\begin{keywords}
Topic coverage, Topic coherence, Topic discovery, 
Topic models, Topic model evaluation, Topic model stability 
\end{keywords}

\titlepgskip=-20pt

\maketitle

\section{Introduction} 

\begin{table*}
\begin{center}
\caption{\footnotesize Examples of interpretable topics of topic models built 
from a dataset of news texts (top) and biological texts (bottom). 
Each topic is characterized by top-weighted topic words. 
Top-weighted topic documents are displayed for the last topic of each dataset . 
Topic labels are the result of human interpretation. 
We note that all the biological topics correspond to concepts of phenotypes (organism \hl{characteristics}).}
{\small
\begin{tabular}{ll} 
\toprule
 Topic label &  Top-10 topic words \\
\midrule
 China & china, chinese, beijing xi, russia, asia, global, region, asian, jinping \\
 Boston Bombing Trial & tsarnaev, boston, bomb, marathon, tamerlan, dzhokhar, trial, penalty, brother, juror \\
 Climate Change & climate, warming, global, scientist, rise, science, scientific, temperature, inhofe \\ 
 & ``Al Gore says climate change deniers should pay `a Price' '' \\
 & ``Smaller percentage of Americans worry about global warming now than in 1989'' \\
 & ``Florida officials say they were banned from saying `Climate Change' \ldots '' \\ 
\midrule
 Spore-forming & spore, sporulate, endospore, germinating, vegetative, survive, coat, forespore \\ 
 Milk Fermentation & dairy, milk, cheese, starter, yogurt, flavor, lactose, ferment, ripen, food \\
Radiat. \& Desicc. Tolerant &
 radiation, repair, desiccation, ionizing, desert, damage, irradiated, radiation-resistant \\ 
 & ``Deinococcus gobiensis: Insights into the Extreme Environmental Adaptations'' \\ 
 & ``Deinococcus maricopensis is an aerobic, radiation-resistant, Gram-positive \ldots '' \\ 
 & ``Deinococcus radiodurans is an extremophilic bacterium, one of the most \ldots '' \\ 
\bottomrule
\label{topic-examples}
\end{tabular}
}
\end{center}
\end{table*}

Topic models \citinl{Blei2003} are unsupervised models that take as input a collection of text documents
and learn a set of topics, constructs represented as weighted lists of words and documents.
A topic of a topic model is expected to be interpretable as a concept, i.e., 
correspond to human understanding of a topic \hl{occurring} in texts.
Examples of interpretable model topics can be found in Table \ref{topic-examples}. 
Topics can help an \hl{analyst} gain insight into textual content, or they can be used 
to create topic-based representations of words and documents for downstream applications.
Since they were introduced, topic models became a popular text analysis and processing tool with numerous applications, 
including \hl{exploratory} text analysis \citinl{Chuang2012}, information retrieval \citinl{Wei2006}, 
natural language processing \citinl{boyd2007topic}, and topic discovery \citinl{Evans2014}.

Although widely used, topic models are prone to random variations and errors due to the stochastic nature of the learning algorithms.
In order to mitigate this problem, a number of topic model evaluation methods has been devised 
\citinl{Wallach2009, Chang2009, DeWaal2008, Newman2010, Chuang2013, Roeder2015, ying2019inferring}. 
These methods aim to provide tools and metrics for the analysis of topic models 
and for the construction of models with interpretable topics.
For example, models can be evaluated using measures of topic coherence \citinl{Newman2010, Roeder2015}, 
or using measures of model stability -- a property of consistent inference of similar topics \citinl{DeWaal2008, Belford2018}.

This paper upgrades an approach to topic model evaluation based on the notion of \emph{topic coverage} \citinl{Chuang2013}, 
i.e., on measuring how well the topics of a topic model cover a set of pre-compiled concepts.
In \citinl{Chuang2013} the authors describe a method for measuring and visualizing 
several types of relations between model topics and a set of concepts defined by domain experts.
The correspondence of topics to concepts is referred to as ``domain relevance'',
and a concept is considered covered if there exists a matching model topic \citinl{Chuang2013}.
The experiments in \citinl{Chuang2013} demonstrate the potential of the coverage approach
for performing automatic analysis of both topic models and measures of model quality,
and show that the relations between concepts and topics depend on models' types and hyperparameters.
Despite the demonstrated potential, there is no follow-up work focused on coverage-based evaluation methods.

Our work approaches topic coverage as a method of quantitative evaluation rooted in the \hl{use case} of topic discovery.
We propose new, reliable, and practical measures of coverage, and perform a series of experiments on two different datasets.
The experiments lead to practical recommendations for topic modeling
and provide insights into both topic models and other methods of model evaluation.
By providing new measures and the first publicly available\footnote{\url{https://github.com/dkorenci/topic\_coverage}}
coverage datasets and tools, we facilitate future research on both topic coverage and novel methods for topic model evaluation.

\newpage
\hl{In summary, our work contributes the following:}
\begin{itemize}
\item \hl{New measures of coverage, including the first unsupervised coverage measure,}
\item \hl{Recommendations for the use of topic models, including the experimental support 
for the use of the} \hla{\nmfmodel} \hl{model} \hla{\citinl{Lee1999},} 
\item \hl{Insights into topic models, including the relationship between 
coverage, the number of model topics, and the size of reference topics,}
\item \hl{Insights into other methods of topic model evaluation, including 
the inability of the standard measures of coherence and stability to detect high-coverage models,}
\item \hl{Coverage datasets and the source code of the measures and the experiments.}
\end{itemize}

The analysis of topic coverage is based on a set of \emph{reference topics} 
and on \emph{measures of coverage} that compute how well the model topics match the reference topics.
Reference topics represent the topics of interest that topic models are expected to discover.
Once a set of reference topics is compiled, \emph{coverage of reference topics by a single topic model instance} 
is the proportion of reference topics covered by model topics. 
A \emph{single reference topic is covered} if there exist \hl{one or more} matching model \hl{topics}.
We use the term ``reference topic'' instead of the term ``reference concept'' used in \citinl{Chuang2013} 
in order to emphasize that a reference topic is a construct represented in the same way as a model topic -- 
as a weighted list of words and documents.

The workflow of coverage-based model evaluation consists of three steps.
In the first step a set of reference topics is constructed. 
In the second step a set of topic models is built, expectedly by varying model types and hyperparameters.
In the third step the measures of coverage are applied to topic model instances and the coverage scores are analyzed.
In the case of using coverage to analyze other measures of model quality, 
these measures are applied to topic models and their scores are correlated with the output of the coverage measures.

The coverage approach \hl{described in this paper} evaluates the models' performance 
in the process \hl{of} topic discovery, a prominent application of topic models.
During the process of topic discovery an analyst examines and interprets the topics of topic models 
in order to find useful topics that can offer insight and be used for subsequent text analysis.
Topic discovery with topic models has been applied, inter alia, in news analysis \citinl{Bonilla2013Dec, Evans2014, Jacobi2016}, 
political science \citinl{Quinn2010, Grimmer2010}, neuroscience \citinl{nielsen2005}, and biology \citinl{Brbic2016}.
Table \ref{topic-examples} contains examples of topics of interest in an analysis of news issues, 
and topics useful for an analyst interested in biological concepts.

From the perspective of topic discovery, coverage achieved by a topic model simply quantifies 
how useful the model would be to an analyst interested in discovering the reference topics.
\hl{In case of an exploratory analysis carried out to obtain a broad topical overview,  
an example set of reference topics would contain high-level topics covering important aspects of texts.
In a more focused analysis, the reference topics would correspond to more specific topics of interest.}
\hl{We note that our approach to coverage is focused primarily on discovering reference topics, 
and that a reference topic might match more than one model topic.
This situation can occur in practice but it does not imply a degradation of models' performance.
In other words, two topic models that cover the same number of reference topics relay
the same amount of useful information to the analyst.}

\hl{In this paper, the} design and evaluation of coverage measures and the coverage experiments are carried out on two datasets.
These datasets represent two different domains for which interest for topic discovery exists: 
journalistic news text and biological text. 
Each of the two datasets consists of a text corpus, a set of reference topics, and a set of topic models.
Each set of reference topics is based on the output of an existing topic discovery study.
The two datasets are described in detail in Section \ref{section.data}.

\hl{The measures of coverage are a basis of a coverage experiment -- their reliability 
determines the reliability of the results, and ease of their construction influences the feasibility of the experiments.}
\hl{Therefore the main} contribution of our work consists of the new measures of topic coverage.
Measure of coverage proposed in \citinl{Chuang2013} matches model topics with reference topics 
via a probabilistic model fitted on data derived from human scores of topic matching.
However, the model and the process of its construction are complex, 
the topic matching scores are crowdsourced from non-experts 
asked to assess similarity of scientific topics, and the measure is not validated.
While the described measure may be used to demonstrate the coverage approach and the related visualization apparatus,
it is hard to reproduce and not suitable for calculation of reliable coverage scores.

We propose a conceptually simple measure of coverage, described in Section \ref{section:supmeasures},
that matches model and reference topics using a standard binary classifier based on a small set of distance-based features.
The classifier is \hl{trained} on a dataset of topic pairs labeled by trained annotators 
acquainted with the topic semantic, and it achieves matching performance close to human agreement. 

Supervised measures of coverage rely on human annotation of topic pairs, 
a time-consuming process that hinders quick application of these measures on new datasets.
Therefore an important contribution of this paper is the 
\hl{measure described in} \hl{Section \ref{section:cdc}}, \hl{which is the first unsupervised measure of coverage.}
It uses topic distance as a criterion for topic matching 
and operates by integrating a range of coverage scores calculated for a range of distances.
We show that this measure has a very high rank correlation with our supervised measure.
The unsupervised measure can be effortlessly deployed on new datasets 
and used for model selection and evaluation by way of ranking a set of topic models.
Furthermore, the measure is based on a curve that is a useful tool for visual analysis and comparison of topic models.

The two proposed measures have applications beyond coverage, 
which we demonstrate in Section \ref{sect:cov4stabil} by adapting them to measure model stability.
The stability measure based on supervised matching provides an experimental support 
for the interpretation of stability as the property of consistent uncovering of the same concepts.
The stability measure based on the adaptation of the unsupervised coverage measure 
correlates almost perfectly with a standard stability measure while being much faster to compute. 

The evaluations of topic models that we perform lead to 
recommendations for the choice of topic models used in topic discovery.
The experiments in Section \ref{sect:modelcoverage}, where we evaluate coverage of topic models of different types, 
show that the \nmfmodel model \citinl{Lee1999, Arora2012} is a good default choice 
for topic discovery due to high coverage, its ability to precisely pinpoint 
the reference topics, and consistent performance on both datasets.
The experiments in Section \ref{sect:covbysize}, where we measure the coverage of reference topics divided into size categories,
support the use of larger models with more topics.
These findings have practical implications since the \ldamodel model \citinl{Blei2003} with a modest number of topics is often a 
default choice for topic discovery.

Our research of the neglected coverage approach also contributes to the broader field of topic model evaluation.
Namely, the amount of work in this field is modest in comparison 
with the amount of research focused on new model architectures. 
On the other hand, there are still no satisfactory methods for 
automatic semantic validation of topic models \citinl{ying2019inferring, doogan2021}.
This hinders the applicability of topic models in computational social sciences \citinl{ying2019inferring}
and for expert analysis of text collections \citinl{doogan2021}.
Popular measures of topic coherence \citinl{Newman2010} are \hl{designed}
to correlate with human coherence scores \citinl{Roeder2015}, 
but it is unclear how well they correlate with models' performance in practice \citinl{ying2019inferring, doogan2021}.
Only recently has an experimental validation of coherence measures been performed \citinl{doogan2021},
and it revealed that these measures are not a reliable guide for model selection \citinl{doogan2021}.
Another approach to automatic model evaluation is based on measures of stability, 
a property of consistent inference of highly similar models.
Stability is claimed to be a \hl{desirable} property of models 
applied in computational social sciences \citinl{Koltcov2014, Chuang2015} or for topic discovery \citinl{Belford2018}.
However, to the best of our knowledge no validation of stability measures has been \hl{performed}.

In \hl{contrast} to the topic coherence and model stability measures, which express abstract 
model qualities, coverage is grounded in the use case of topic discovery
and the coverage scores are interpretable in terms of a match with ground truth reference topics
-- a set of interpretable topics of interest to an analyst.
On the other hand, coverage-based evaluation relies on a fixed topic modeling scenario, 
defined by a text collection (model input) and a set of reference topics (model output).
Therefore coverage cannot be applied for model selection in future applications,
but rather for large scale automatic analysis of both topic models and measures of model quality.
While the measures, experiments, and datasets we contribute are a starting point, 
the findings based on analysis of coverage will become more robust 
as new datasets representing new application settings are constructed and made available.

Experiments in this paper provide new data points that improve 
the understanding of both topic models and measures of models quality.
In Section \ref{sect:covbysize} we measure the coverage of reference topics divided into size categories.
The results show that small models with fewer topics can cover only
large (frequently \hl{occurring}) reference topics, while the larger models are able to uncover topics of all sizes. 
The experiment described in Section \ref{sect:covbytype} is motivated by the use case of news topic discovery in social sciences.
We measure coverage of news topics categorized as either corresponding to a news issue or not, 
and as being either abstract or concrete.
The experiment demonstrates that semantic categories of topics influence their coverage by topic models.

In Section \ref{sect:covcohstabil} we apply the measures of coverage to analyze 
the measures of topic coherence \citinl{Newman2010} and model stability \citinl{DeWaal2008}.
Comparison of coverage and topic coherence, performed in Section \ref{sect:covcoh}, 
shows that no strong and consistent correlation between the two exists.
These results are consistent both with a prior comparison of coverage and coherence \citinl{Chuang2013}, 
and with a recent study that validates coherence measures \citinl{doogan2021}.
The experiments in Section \ref{sect:covstabil}, examining the relation between coverage and model stability, 
are to the best of our knowledge the first attempt to semantically analyze measures of stability.
The experiments show no correlation between the two, demonstrating that model stability does not necessarily imply model quality.

\section{Datasets} 
\label{section.data}

We perform the coverage experiments on two distinct text domains -- news text and biological text.
For each text domain we construct a dataset, which consists of three components:
a text corpus, a set of reference topics, and a set of topic models.
Such a dataset is the basis for coverage experiments and the construction of coverage measures.
We refer to these two datasets as the \emph{news dataset} and the \emph{biological dataset}.

The two datasets represent two different text genres: journalistic texts describing 
political news and expert biological \hl{texts} describing microorganisms.
The datasets are based on text corpora and reference topics from two \hl{previous} experiments
in which topic models \hl{were} used for topic discovery on news \citinl{Korencic2015} and biological \citinl{Brbic2016} text.
The reference topics therefore represent useful output of topic discovery, 
while being representative of concepts discoverable by standard topic models.
The set of topic models built for each dataset contains models of standard types,
configured with varying number of topics.
We next describe the three components of the two datasets.

\subsection{Text Corpora}
\label{section:corpora}

A text corpus is a basis of a dataset since both the reference topics and the topic models are derived from the corpus texts. 
The dictionary and the document index associated with a corpus
are a basis for representing the reference and model topics as topic-word and topic-document vectors.

\paragraph{News corpus}
The news dataset is based on the corpus of mainstream US political news 
collected by \citnoun{Korencic2015} for evaluating topic model approaches to news agenda analysis.
The texts were collected from popular news sites during a three-month period, after which filtering of non-news 
texts and deduplication was performed, resulting in a total of 24.532 texts.
Topic modeling was \hl{preceded} by text preprocessing that consisted of stop-word removal, morphological normalisation, 
and removal of low- and high-frequency words. The final dictionary contains 23.155 words.

\paragraph{Biological corpus}
The basis of the biological dataset is the corpus of texts about bacteria and 
archea microorganisms used for the discovery of phenotype topics \citinl{Brbic2016}.
This corpus contains texts about 1.640 distinct species obtained from five sources:
the Wikipedia, the MicrobeWiki containing texts about microorganisms, the HAMAP proteomes database containing protein-related microorganism data, 
the PubMed database of paper abstracts, and the PubMed Central database of full-text papers.
The final corpus contains 5994 documents.
The documents were preprocessed by removing English stop-words, parts of the texts containing references,
and the words with frequency less than four, after which the words were stemmed.
The final dictionary used in coverage experiments contains 6259 words that ocurr in at least 4 of the original text sources.

\subsection{Reference topics}
\label{section:reftopics}

The set of reference topics defines the measured coverage -- 
by definition, topic models with high coverage are the models capable of detecting a large \hl{proportion} of reference topics.
We conduct our experiments with two sets of reference topics -- the news and the biological reference topics.
Each set of reference topics is based on an output of a topic discovery study, i.e.,
obtained by human inspection and interpretation of topic models' topics.
In other words, these reference topics are an interpretable and error-free output of topic models,
and represent useful concepts discoverable by model topics.

News reference topics are a result of an analysis of the media agenda \citinl{Korencic2015} performed with topic models built form news articles.
\hl{In other words, the news topics represent a broad range of topics that occurr in the news.}
These topics \hl{correspond to} persons, organizations, events and stories, 
and abstract concepts such as news issues and topics related to economy and politics.
Biological reference topics are a result of topic discovery performed on biological texts describing microorganisms \citinl{Brbic2016}, 
and correspond to concepts of phenotypes, organism characteristics, such as termophilia and various types of pathogenicity.

Using a set of reference topics that the topic models are able to discover is a decision made to ensure that
models' coverage will not be skewed due to the nature of reference topics.
Namely, we wish to avoid the scenario where the models display low performance and are mutually indistinguishable because 
the reference topics represent a subset of topics hard to cover.
Therefore we alleviate the coverage problem and leave the harder cases of reference topics for future research.
We hypothesize that examples of such hard cases are very specific concepts and high-level abstract concepts devised by humans.
Despite being within the models' reach, the reference topics used in the experiments are not trivial to uncover.
The subsequent coverage analysis shows that the models cover at best $64\%$ of the reference topics 
in the case of news topics, while for the biological topics the best case coverage is $44\%$.

From the machine perspective, a reference topic is represented in the same way as any model topic --
with a vector of topic-word weights and vector of topic-document weights.
The components of these vectors correspond, respectively, to the words in the 
corpus dictionary and to the indices of the corpus documents.
This way all the topic-related computations \hl{necessary} for the calculation of coverage,  
such as calculation of distance between topics, make no distinction between the reference topics and the model topics.

Each set of reference topics is constructed in three main steps. 
The first \hl{step} consists of building the topic models. 
In the second step the models' topics are inspected, interpreted, and filtered. 
Only uninterpretable news topics are filtered out and no concept-type restrictions are imposed, 
while all the biological topics that do not correspond to phenotypes are filtered out.
Finally, the topic-word and topic-document vectors of the reference topics are constructed from the corresponding model topics.
Each of the two methods of reference topics construction reflects the specifics of 
the corresponding topic discovery approach \citinl{Korencic2015, Brbic2016}.
The details of the methods are described in Appendix \ref{app:reftopics}.

\subsection{Topic Models}
\label{section:covmodels}

There exists a large number of topic model types representing 
a range of assumptions and approaches to modeling text structure. 
We apply the proposed coverage methods to evaluate models from two standard categories
-- probabilistic topic models \citinl{Blei2012} and topic models based on non-negative matrix factorization \citinl{Lee1999, Arora2012}.
There exist numerous model variants belonging to these two categories,
as well as alternative architectural approaches such as 
geometric \citinl{yurochkin2017} and neural \citinl{zhao2021topic} topic models.
However, the structure of a topic model is not an apriori guarantee for model performance \citinl{grimmer2013text}, 
and each model should be validated within the context of its application \citinl{grimmer2013text}.
Therefore we focus on the evaluation of topic models commonly used in topic discovery
and leave the evaluation of many other models for future work.

In the coverage experiments that follow we evaluate models from the categories 
of parametric and nonparametric probabilistic models, and a model based on non-negative matrix factorization.
The evaluated models all make minimal assumptions about the structure of text, 
similar to the assumptions of the seminal \ldamodel model \citinl{Blei2003}. 
This makes these models applicable to a generic use case of topic discovery performed on a collection of text documents.
Specifically, each of the models assumes that the text of a document can be 
approximated with a weighted mixture of topics, where each topic is a weighted list of words.
Other text-related variables such as sentiment and various metadata \citinl{Blei2012} are not included in the models' structure.

Regardless of type, each model is represented simply as a set of topics,
and each topic is represented with vectors of topic-word and topic-document weights.
As noted earlier, the reference topics are represented in the same way.
This black-box view of topic models makes 
the proposed coverage methods applicable to a wide variety of topic model types.

First of the model types we experiment with is the seminal Latent Dirichlet Allocation model \ldamodel \citinl{Blei2003}.
The \ldamodel model is representative for a wide variety of model types, many of which are its direct extension.
\ldamodel assumes a fixed number of topics, and the topic-word and topic-document relations
are modeled with matrices of word-in-topic and topic-in-document probabilities. 
Probabilistic inference algorithms, such as variational inference \citinl{Blei2003} and Gibbs sampling \citinl{Griffiths2004},
are capable of learning the topic data from a set of unlabeled text documents.
The \ldamodel model has been widely applied in many topic modeling tasks, including topic discovery 
\citinl{Griffiths2004, dimaggio2013, Evans2014, Puschmann2016, Jacobi2016, Maier2018}.

The second model type is a modification of the \ldamodel model 
to which we will refer to as ``asymmetric LDA'' (\aldamodel).
The \ldamodel model assumes that the prior for the document-topic distribution is symmetric, 
which means that all the topics have an equal prior probability of appearing in a document.
In contrast, the \aldamodel model allows for an asymmetric prior learnable from data, 
implying topics with varying prior probabilities.
This allows for more flexibility in modeling of the document-topic relation
and in effect allows for the topics to be recognized, on the level of the text collection, as being larger or smaller.
This \hl{approach} potentially leads to higher topic quality \citinl{Wallach2009priors} and better detection of smaller topics \citinl{Wang2015}.
The \aldamodel variant we experiment with is implemented in the HCA software package \citinl{Buntine2014},
by way of using normalized Gamma priors to model the document-topic distribution.

The third model type is a nonparametric topic model based on Pitman-Yor priors \citinl{Buntine2014}, denoted \pypmodel. 
Unlike the other models, the \pypmodel model is able to learn the number of topics from data.
The \pypmodel model, denoted NP-LDA in \citinl{Buntine2014}, is an extension of the nonparametric 
HDP topic model based on Hierachical Dirichlet Process \citinl{Teh2006}.
The HDP model generalizes the \ldamodel model by using a probability distribution over a countably infinite collection of topics \citinl{Teh2006}.
The \pypmodel model generalizes the HDP model by using the more flexible Pitman-Yor process \citinl{Pitman1997} 
to model a distribution over the infinite set of topics. 
The nonparametric models have been applied for topic discovery \citinl{kim2014computational} and it is our intuition
that the added flexibility of learning the number of topics might lead to better coverage, especially coverage of the smaller topics.

The fourth topic model type, denoted \nmfmodel, utilizes non-negative matrix factorization \citinl{Lee1999, Arora2012}.
The \nmfmodel model is based on approximation of the text collection, represented as matrix of document-word weights, 
in terms of a product of non-negative matrices containing document-topic and topic-word weights.
In other words, the topics are the latent factors optimized to approximate 
the original text matrix under the assumption of non-negativity.
The \nmfmodel model has been \hl{successfully} used for topic discovery in several scenarios \citinl{nielsen2005, Choo2013, Greene2015, Brbic2016}, 
and has the potential to produce topics with quality equal to or better then the quality of \ldamodel topics \citinl{ocallaghan2015}.

Finally, we describe the set of topic model instances used in the coverage experiments. 
For each of the model types, with the exception of the nonparametric \pypmodel model, 
the parameter $T$ defining the number of model topics is varied, 
as this important parameter defines a model's capacity and affects the structure of its topics.
Namely, $T$ influences topic granularity \citinl{Chang2009, Greene2014, Evans2014, blair2016increasing},
in such a way that a small $T$ leads to broad and general topics, while a large $T$ results in fine-grained and specific topics.
The news and the biological datasets contain, respectively, $133$ and $112$ reference topics. 
We build the topic models by varying $T$ between values of $50$, $100$, and $200$.
These choices of $T$ correspond, respectively, to a number of topics that is smaller then, 
approximately equal, and larger than the number of reference topics.
For the nonparametric \pypmodel model the maximum number of learnable topics is set to $300$.

For each combination of the model type and the number of topics,
$10$ model instances are built with different random seeds in order 
to account for \hl{stochastic} variation, i.e., to obtain a more robust assessment of coverage.
Therefore, for each dataset a total of $100$ model instances are built: 
$10$ instances for each pair of the model type (\ldamodel, \aldamodel, or \nmfmodel) and the number of topics ($50$, $100$, or $200$), 
plus additional $10$ instances of the nonparametric \pypmodel model.

For each of the two datasets the topic model instances are inferred from the texts of the corresponding corpus, 
preprocessed by performing stopword removal and word normalization.
The text corpora and the preprocessing methods are described in Section \ref{section:corpora}.
Appendix \ref{app:modelbuild} contains the details of topic model construction
that include the choice of hyperparameters, learning algorithms, and software tools.

\section{Measures of Topic Coverage}
\label{section:measures}

Measures of topic coverage compute scores that quantify how well the topics of a topic \hl{model} cover a set of reference topics.
In this section we propose two distinct measures of topic coverage -- the \sups measure, 
based on supervised approximation of human intuition of topic matching, 
and an unsupervised \cdc measure, designed to approximate 
the supervised measure and serve as a quickly deployable model selection tool.
The \cdc measure calculates coverage by using a distance threshold as a topic matching criterion 
and aggregates the coverages obtained by varying the threshold.

\hl{The measures we propose are the first coverage measures that
are validated and straightforward to construct, while the} \hla{\cdc} 
\hl{measure is the first unsupervised measure of coverage.}
\hl{These measures are the main contribution of this paper 
since they improve both the reliability and the feasibility of the coverage experiments.}

\subsection{Coverage Based on Supervised Topic Matching}

\label{section:supmeasures}

The supervised topic coverage measure mimics the procedure 
in which a human annotator \hl{assesses} weather there exists a model topic \hl{that matches} a reference topic.
Model coverage can be calculated from this matching information
as a proportion of reference topics matched by at least one model topic.

Although this procedure would result in coverage scores based on human knowledge, 
it is time-consuming and impractical, especially for large sets of topic model instances.
We solve this problem by constructing a supervised model that approximates human intuition of topic matching. 
Once such a model is available, it can be used for automatic calculation of coverage of arbitrarily many topic models.

We base our solution on a dataset of topic pairs labeled with matching scores of human annotators. 
Maching of two topics is defined as the equality of concepts \hl{obtained} by the topics' interpretation.
The topic matching problem is cast as a problem of binary classification 
of topic pairs into \emph{matching} and \emph{not-matching} classes.
Several standard classification models are constructed and evaluated, 
and the best performing model is used in the subsequent experiments for coverage score calculation.
The process of data annotation and model building is performed for both the news and the biological dataset.

\hl{
In order to calculate the supervised coverage of a set of $R$ reference topics 
by a topic model with $T$ topics, every reference topic has to be matched, in the worst case, with every model topic.
The matching operation consists of feature construction and of the computation of the classifier's output.
Our features consist of distances between topic-word and topic-document vectors,
which can be calculated in time proportional to either the vocabulary size $V$ or to the number of documents $D$.
The application of the classification model to the features requires constant time.
Therefore the asymptotic complexity of calculating supervised coverage is $\mathcal{O}(R T (V+D))$.
}

\subsubsection{Topic Pairs Dataset}
\label{sect:topicpairdataset}

In order to learn a matching model that generalizes well to different types of topics,
the dataset of topic pairs is sampled from both the reference topics and the topics of topic models of different types and sizes. 

However, in a randomly sampled set of topic pairs a large majority of pairs consists of non-matching topics.
This means that supervised topic matching is an imbalanced learning problem \citinl{Branco2016, Krawczyk2016},
a scenario in which inference of high-performing models is hindered since only a small fraction 
of learning examples that define the structure of the positive class is available.
In our case, the positive examples are pairs of matching topics. 

Solutions to this problem include active learning and resampling methods \citinl{Branco2016}, 
but we opt for a simpler solution applicable to topic pairs.
This solution relies on the intuition that mutual distance of 
two topics is in an inverse correlation with the probability of topics' semantic match.
Concretely, we sample topic pairs according to mutual distance of topics in order 
to \hl{achieve} a higher proportion of pairs with mutually close topics that have higher probability of matching.
An inspection of a validation sample of topic pairs confirms that the described procedure leads to a balanced dataset.
The elaboration of the problem and the details of the solution can be found in Appendix \ref{app:pairbalance}.

The final dataset of representative topic pairs used for model construction is created in the following way. 
First, a large set of topics is created by building one model instance 
for each combination of model type and number of topics and taking all the topics of the chosen instances. 
Next, three copies of each of the reference topics are added to the topic set
in order to make the number of reference topics approximately equal to the number of topics of each of the model types.
After that a set of all the distinct pairs of two different topics is created 
and these pairs are divided into ten subsets corresponding to equidistant intervals of topics' cosine distance.
Finally, $50$ pairs are sampled randomly from each of the subsets, leading to a total of $500$ topic pairs.
Of these $500$ pairs, $300$ pairs labeled by the annotators are used for model learning, 
while the remaining pairs are used for training and calibration of the human annotators.

\subsubsection{Annotation of Topic Pairs}
\label{sect:pairlabeling}

Next we describe the procedure used to annotate topics pairs with human scores of topic matching.
The key aspect, which defines the nature of the \hl{matching} approximated by the supervised model, is the definition of a topic match.
We define a topic match as conceptual quality of topics --
two topics are considered equal if they are interpretable as the same concepts, 
where the interpretation of a topic as a concept is as specific as possible.
This definition is in line with the approach of measuring precise coverage of the reference topics --
we wish to assign high scores to models with topics that match the reference topics precisely.
The alternative to this approach would be to focus 
on matching similar topics, such as sub-topics, super-topics, and \hl{overlapping} topics.

\begin{table*}
\caption{\footnotesize Performance of human annotators and supervised models on the task of matching topic pairs. 
Both the average scores and the standard deviations are displayed.}
\centering
\normalsize
{\small
\begin{tabular}{lcccc}
\toprule
 & \multicolumn{2}{c}{News dataset} & \multicolumn{2}{c}{Biological dataset} \\
\midrule
& \fone & st.~dev. & \fone & st.~dev. \\
\emph{human annotators} & $0.854$ & $0.034$ & $0.817$ & $0.032$ \\
Logistic regression & $\mathbf{0.835}$ & $0.079$ & $\mathbf{0.787}$ & $0.080$ \\
Support vector machine & $0.830$ & $0.077$ & $0.784$ & $0.080$ \\
Random forest & $0.804$ & $0.110$ & $0.776$ & $0.057$ \\ 
Multilayer perceptron & $0.814$ & $0.106$ &  $0.723$ & $0.062$ \\
\bottomrule
\end{tabular}}
\label{table:supcovresults}
\end{table*}


More precisely, two topics -- constructs described by weighted lists of words and documents --
can differ both semantically, on the level of interpreted concepts, 
and due to the noise caused by stochastic topic model learning algorithms.
This random variations manifest as a certain proportion of random or unrelated words and documents within the topic.
On the semantic level, we define topic equality as matching of
concepts obtained by interpreting topics as specifically as possible.
Matching of concepts is defined as equality or near equality of concepts, allowing small variations and similar aspects of a same concept.
Stochastic differences are accounted for by labeling topics as equal but with presence of noise.
This is the case when one or both topics contain a noticeable amount of noise but the topics are still
interpretable and the equality of interpreted concepts exists as \hl{previously} defined.

Based on the previous definition, a pair of topics is labeled with $1$ in case of 
topic equality, i.e., when concepts match without noise. 
A pair is labeled with $0.5$ in case of a match with the presence  
of noise or small semantic variation, and with $0$ when the concepts do not match.
Preliminary experiments showed that such labeling is simpler and more consensual for annotators 
than labeling on the binary scale that accounts only for the \hl{possibilities} of match and mismatch.

In order to ensure the quality and consistency of annotations, the annotation was conducted
according to the methodology of content analysis \citinl{Krippendorff2012} --
precise instructions were provided, the annotators had the knowledge required 
to interpret the texts and the topics, and were trained
until the measure of mutual agreement reached a satisfactory level.

The annotation process resulted in a set of $300$ topic pairs, each annotated by three annotators.
The annotated pairs serve as training data in the process of building the supervised model of topic matching.
The details of the annotations process are described in Appendix \ref{app:pairannot}.

\subsubsection{Supervised Topic Matcher}
\label{sect:supmodelconstr}

At the heart of the proposed supervised measure of topic coverage is a
binary classification model that approximates, for a pair of topics, human assessment of weather the topics match or not.
This model is used to compute weather reference topics are covered by topics learned by a topic model.
In this section we describe the method of construction of such a classification model.

The classification problem is defined in the following way.
Each topic pair was annotated by three annotators with one of the three possible labels.
The matching labels are $1$ -- concepts obtained by topic interpretation match, 
$0.5$ -- topics match but either certain amount of noise 
or a small semantic variation exist, and $0$ -- no match.
Binary labels are obtained by averaging the labels and applying the decision threshold of
$0.75$ -- if the label average is above $0.75$ the topic pair is assigned the positive class that designates a match, and negative class otherwise.
In other words, topics are labeled as matching if at least two of the annotators labeled the pair as matching
while the third annotator labeling the pair as at least partially matching.
The averaging of the annotators' scores is performed in order to obtain more robust labels
and the definition of the positive and negative class corresponds to matching of topics on a precise level. 

We consider four standard classification models: logistic regression \citinl{Murphy2012}, support vector machine \citinl{Cortes1995}
with radial basis function kernel, random forest \citinl{Breiman2001}, and multilayer perceptron \citinl{Murphy2012}.

Topic pairs are represented as features based on four measures of distance: 
cosine distance, Hellinger distance \citinl{Jebara2003}, \loned distance, and \ltwod distance.
These four distance measures are applied to both the pair of normalized topic-word vectors
and to the pair of normalized topic-document vectors.
Thus the input for classification of a topic pair consists of eight distance-based features, 
four based on topic-related words and four based on topic-related documents.

For each classification model, hyperparameter optimization is performed 
using nested five-fold crossvalidation in combination with the \fone measure as the performance metric.
Nested five-fold crossvalidation is used to obtain robust \hl{assessment} of the quality 
of the model variant with optimized hyperparameters \citinl{Cawley2010}.
The optimization is \hl{performed} on the entire dataset of $300$ labeled topic pairs.
The details of feature construction and model construction are described in Appendix \ref{app:supmodels}.

The performances of the optimized classification models, for both datasets, are laid out in Table \ref{table:supcovresults}.
For each model and dataset, the table contains both the average and the standard deviation of 
\fone calculated on five outer folds of nested five-fold crossvalidation.
Model performance data is complemented with the scores of mutual agreement of human annotators, also measured by \fone.
The human agreement is calculated as average \fone score of a single annotator 
predicting the class labels obtained by averaging the annotations of the two remaining annotators. 
Specifically, annotators' binary class labels are calculated by averaging 
the annotators' scores and applying the $0.75$ threshold to decide if the topics in a pair match. 

\begin{figure*}[h]
\centering
\includegraphics[width=2\columnwidth]{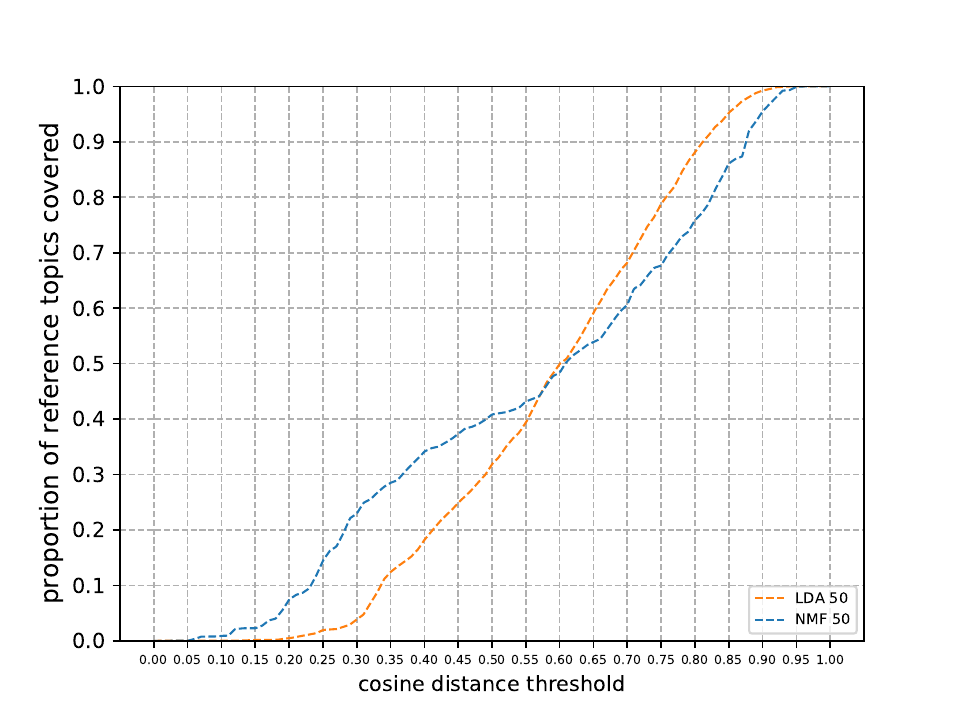}
\caption{\footnotesize Coverage-distance curves depicting coverage of reference topics by the \ldamodel and \nmfmodel models with 50 topics, 
for the collection \hl{of news} texts.}
\label{fig:cdcexample}
\end{figure*}

Table \ref{table:supcovresults} shows that the logistic regression model
has the highest \fone values on both datasets.
The support vector machine model is a close second, while the other two models, 
multilayer perceptron and random forest, are not far behind. 
In addition to logistic regression having top \fone scores, 
it is structurally the simplest model with the smallest number of hyperparameters, 
which we view as an additional advantage. 
We therefore choose the logistic regression model as the basis of the supervised coverage measure,
i.e., as the model for matching reference topics with model topics in order to calculate coverage of the reference set.
The final models used to measure coverage in the following experiments are obtained, for each dataset, 
by first optimizing the hyperparameters with five-fold crossvalidation 
and then learning the final model with optimized hyperparameters.
Both hyperparameter optimization and model learning are performed on the entire set of $300$ labeled topic pairs.

The classification results show that the performance of the supervised models is close to the mutual agreement of human annotators.
This shows that the described process leads to supervised models 
that can approximate human assessment of topic matching well.
Specifically, the human scores of topic matching, based on the equality of the interpreted concepts,
can be approximated well from a small set of features based on distances between topic-word and topic-document vectors.

This finding is applicable wherever there is a need for
automatic matching of topic-like constructs defined by weighted words and documents.
In the context of application of supervised topic matcher for coverage calculation, 
the previous results support the claim that the computed coverage will be reliable and close to human assessment.

\subsection{Coverage-Distance Curve} 
\label{section:cdc}

The construction of the supervised measure of coverage requires a time-consuming process of construction of a labeled dataset of topic pairs.
This process includes both recruiting and training of annotators with sufficient knowledge of the text domain, and the process of topic annotation.
\hl{Conversely}, an unsupervised measure of coverage could be quickly applied to a new evaluation scenario.
We propose such a measure and show that it correlates very well with the supervised measure \sups, 
which makes it applicable for ranking and selection of topic models by coverage.

The unuspervised measure, similarly to supervised coverage, computes coverage of a set of reference topics by matching them to the model topics.
However, the decision weather two topics match is based simply on a measure of topic distance and a distance threshold
-- two topics match if their distance is below a threshold.
For a specific threshold, the coverage of reference set is the proportion 
of reference topics for which a matching model topic exists.
In order to render the measure threshold invariant, the final coverage 
score is calculated by varying the distance threshold and integrating 
all the coverages corresponding to different threshold values.
More precisely, by varying the distance threshold and calculating corresponding coverage values
a curve is formed, composed from points with x-coordinates corresponding to thresholds and y-coordinates corresponding to coverages. 
In other words, this curve is a graph of a function that maps distance thresholds to corresponding coverages.
We call this curve the Coverage-distance curve and refer to it as \cdcurve{}.
The final coverage measure, which we label as \cdc, is then calculated as the area under the \cdcurve{}. 

The \cdcurve{} illustrates the \hl{dependence} of coverage on the topic distance used as a criterion of topic match.
It can therefore be used as a tool for graphical analysis of the coverage of a 
single topic model and for coverage-based comparison of a number of models.
Figure \ref{fig:cdcexample} contains \cdcurve{s} depicting coverages of 
news reference topics by \ldamodel and \nmfmodel models with $50$ topics. 
It can be seen that for the cosine distance threshold of $0.4$, coverages for \ldamodel and \nmfmodel
are approximately $20\%$ and $35\%$, respectively. 
This means that if two topics are considered equal when their cosine distance is $0.4$ or smaller, 
$20\%$ of reference topics are matched by at least one \ldamodel model topic, 
with this percentage being $35\%$ in case of the \nmfmodel model. 
Inversely, the curve can be used to determine the distance threshold, i.e., 
the required precision of topic matching, necessary to achieve certain level of coverage.
The corresponding values of the \cdc coverage score are $0.410$ for the \ldamodel model and $0.434$ for the \nmfmodel model.
However, the curves illustrate finer differences in the nature of coverage. 
Concretely, the \nmfmodel model has better coverage for smaller distance thresholds
while the \ldamodel model has better coverage for larger thresholds.
This means that the \nmfmodel topics match the reference topics more precisely
and give better coverage under the assumption of stricter criteria of topic match.
On the other hand, the \ldamodel model has better coverage when topic match is more approximate
and a reference topic is allowed to be matched by a model topic at a lower degree of similarity.
This example also illustrates the intuition behind the \cdc score -- 
a model with a higher score is expected to have a more elevated \cdcurve{}
\hl{than} a model with a lower score, which means that it covers more reference topics at lower distance thresholds.

\cdcurve{} and \cdc measure have both similarities and differences with the popular
receiver operating characteristics (ROC) curve and the associated 
Area Under the ROC Curve (AUC) metric \citinl{Ling2003} applicable for evaluation of predictive machine learning models.
The difference is that the AUC measures the \hl{performance} of a set of related predictive models,
usually instances obtained by varying an important model parameter.
The points on the ROC Curve describe model instances -- 
each point has coordinates corresponding to model sensitivity and model specificity.
On the other hand, \cdc is a measure of coverage of reference topics
calculated for a single instance of an unsupervised topic model, 
and each point on the \cdcurve{} has coordinates corresponding to a distance threshold and the derived coverage. 
However, both methods build a curve that provides information about model \hl{behavior} in different scenarios -- 
the ROC curve illustrates the sensitivity/specificity trade-off 
while the \cdcurve{} illustrates the \hl{dependency} between matching distance and coverage.
And each of the two measures is calculated as the area under the corresponding curve, i.e.,
by integrating model performance over a range of options.

\begin{table*}[h]
\caption{\footnotesize Spearman and Pearson correlations between the \cdc measure and both the \sups measure and its variant without cosine features.
For each correlation \hl{coefficient}, a $95\%$ bootstrap confidence interval is shown.} 
\label{table:covcorr}
\centering
\normalsize
\begin{tabular}{lllll}
\toprule
             & \multicolumn{2}{c}{News dataset} & \multicolumn{2}{c}{Biological dataset} \\            
\cmidrule(lr){2-3} \cmidrule(lr){4-5}
             & Spearman     & Pearson      & Spearman     & Pearson      \\
\midrule         
\sups    & 0.97  \sgmnt{0.95}{0.98}   & 0.96 \sgmnt{0.95}{0.97} & 0.95   \sgmnt{0.90}{0.97} & 0.95 \sgmnt{0.94}{0.96}  \\
\sups-nocos  & 0.96 \sgmnt{0.94}{0.97} & 0.93  \sgmnt{0.91}{0.95}   & 0.95  \sgmnt{0.90}{0.96} & 0.96  \sgmnt{0.95}{0.97}    \\
\bottomrule
\end{tabular}
\end{table*}

\cdc and the \cdcurve{} are based on a measure of distance between two topics that serves as a criterion of topic match. 
This \emph{base distance measure} should satisfy three criteria. 
First, a base measure of distance should be bounded, i.e., restricted to finite range of values. 
This property is necessary because the \cdcurve{} is constructed by varying the distance threshold from minimum to maximum distance.
A bounded distance measure also enables a comparison between two different models, 
since their corresponding curves will be constructed over the same threshold range.
Second, the measured distance between topics should correlate well with human intuition of topic similarity. 
Concretely, smaller distances between two topics should correspond to higher probability of topic match, and vice versa.
While this requirement is a reasonable \hl{guideline}, the final test of the measure's semantic 
is the comparison between the \cdc measure and the supervised coverage based on human annotations.
Finally, in order for the \cdc measure and the \cdcurve{} to enable comparison between models of different types,
the semantic of the base measure should be insensitive to the model type.
This requirement is best exemplified by considering the \nmfmodel model that produces topics with 
unbounded positive values and the probabilistic models with topics that are probability distributions.
Topic-word vectors of \nmfmodel topics can thus contain much larger values, 
which can affect coordinate distance measures such as the \loned distance. 
The \loned distance between an \nmfmodel topic and an \ldamodel topic is therefore expected to be larger 
than a distance between two \ldamodel topics, regardless of the semantic similarity of topics.
Intuitively, a sensible base distance measure should be based on relative 
proportions of topic-word weights, which is more similar to human approach to topic matching.

We opt to use, based on the previous considerations, cosine distance between topic-word 
vectors as the base measure of distance between two topics.
Namely, topic-word vector is a standard representation of model topics commonly used 
for calculating topic distance or similarity \citinl{Zhao2011, Chuang2013, Roberts2015, Chuang2015}.
Likewise, cosine distance is a standard measure widely used in text mining for comparing high-dimensional vectors \citinl{Tan2005} 
and experiments with topic models show that it correlates well with human intuition of topic similarity \citinl{Chuang2013}.
For two vectors $v$ and $w$, the cosine distance is defined as: $\cosd(v, w) = 1 - \frac{v \cdot w}{\|v\|\|w\|}$.
The cosine distance is an inverse of cosine similarity of $v$ and $w$, defined as $\frac{v \cdot w}{\|v\|\|w\|}$,
and corresponding to the cosine of the angle formed by vectors $v$ and $w$.
By definition, the cosine distance is bounded and takes on values between $0$ and $2$.
In case of the large majority of topic models that have non-negative topic vectors, 
such as probabilistic topic models and non-negative matrix factorization models, 
the cosine distance takes on values between $0$ and $1$.
Cosine distance thus satisfies all of the previous criteria. 
It is bounded by definition and expected to correlate well with human intuition of semantic distance.
Since it is based on an angle between topic vectors it is also invariant to the sizes of these vectors, 
i.e., the absolute values of topic-word weights that can vary depending on model type.

The proposed \cdc measure is envisioned as a good approximation of the supervised coverage measure \sups 
that can be quickly deployed for selection of high-coverage models.
We thus evaluate the \cdc measure, based on cosine distance of topic-word vectors, 
by calculating its Spearman rank correlation \hl{coefficient} with the \sups measure on the level of topic model.
This correlation shows how well the \cdc-induced ordering of topic models approximates the ordering induced by \sups coverage.
Standard Pearson coefficients of linear correlation are also calculated in order to get a more complete picture of the measure's properties.
For each dataset, the correlations are calculated on the set of $100$ topic models 
of different types and sizes described in section \ref{section:covmodels}.
The $95\%$ bootstrap confidence intervals of the correlation coefficients 
are calculated using the percentile method and $20.000$ bootstrap samples.
We note that both the \cdc and the \sups measure use cosine distance 
-- supervised topic matcher uses features based on cosine distance, 
specifically cosine distances of topic-word and topic-document vectors.
To check if this influences the \hl{strength} of correlation, we built 
a supervised matching model that does not use features based on cosine distance
and calculated correlations between the \cdc and the supervised coverage based on 
this model, which we denote \sups-nocos.
Except for the difference in features, the supervised matching models are the same --
based on logistic regression with hyperparameters optimized by five-fold crossvalidation, as described in Section \ref{section:supmeasures}.

\begin{table*}[h!]
\caption{\footnotesize Coverage of reference topics by topic models of various types and sizes, measured by the \sups and \cdc measures.
For each measure and dataset, the \bst{best score} and the \good{best scores for each number of topics} are indicated.
For each coverage score, a $95\%$ bootstrap confidence interval is shown.}
\label{table:coverage}
\centering
\normalsize
\begin{tabular}{lllll}
\toprule
 & \multicolumn{2}{c}{News dataset} & \multicolumn{2}{c}{Biological dataset}  \\
\cmidrule(lr){2-3} \cmidrule(lr){4-5} 
        & \sups   & \cdc &  \sups &  \cdc  \\
\midrule      
\ldamodel-50   & 0.14 \sgmnt{0.13}{0.15} & 0.41 \sgmnt{0.41}{0.41} & 0.01 \sgmnt{0.00}{0.01} & 0.31 \sgmnt{0.30}{0.31}   \\
\ldamodel-100  & 0.31 \sgmnt{0.30}{0.33}  & 0.51 \sgmnt{0.50}{0.51}   & 0.07  \sgmnt{0.06}{0.08} & 0.40  \sgmnt{0.40}{0.41}  \\
\ldamodel-200  & 0.47 \sgmnt{0.46}{0.48} & 0.60 \sgmnt{0.59}{0.60}   & 0.16 \sgmnt{0.15}{0.17}  & 0.50 \sgmnt{0.50}{0.51}   \\
\aldamodel-50  & 0.12 \sgmnt{0.10}{0.13}  & 0.40  \sgmnt{0.40}{0.41}  & 0.01 \sgmnt{0.01}{0.01}  & 0.31  \sgmnt{0.31}{0.32}  \\
\aldamodel-100 & 0.27 \sgmnt{0.25}{0.28}  & 0.51 \sgmnt{0.51}{0.51} & 0.07 \sgmnt{0.06}{0.08}  & 0.41 \sgmnt{0.41}{0.42} \\
\aldamodel-200 & 0.42 \sgmnt{0.40}{0.44}  & 0.60 \sgmnt{0.60}{0.61} & 0.15 \sgmnt{0.14}{0.16}  & 0.52 \sgmnt{0.51}{0.52} \\
\nmfmodel-50   & \good{0.22} \sgmnt{0.22}{0.23} & \good{0.43} \sgmnt{0.43}{0.44} & \good{0.11} \sgmnt{0.11}{0.12} & \good{0.39} \sgmnt{0.39}{0.40} \\
\nmfmodel-100  & \good{0.40} \sgmnt{0.39}{0.41} & \good{0.56} \sgmnt{0.56}{0.56} & \good{0.22} \sgmnt{0.21}{0.23} & \good{0.54} \sgmnt{0.53}{0.54} \\
\nmfmodel-200  & \good{0.54} \sgmnt{0.53}{0.55} & \good{\bst{0.65}} \sgmnt{0.64}{0.65} & \bst{0.44} \sgmnt{0.43}{0.45} & \bst{0.67} \sgmnt{0.67}{0.68} \\
\pypmodel  & \bst{0.64} \sgmnt{0.62}{0.65} & \bst{0.65} \sgmnt{0.64}{0.65} & 0.23 \sgmnt{0.22}{0.24}  & 0.56 \sgmnt{0.56}{0.56} \\
\bottomrule
\end{tabular}
\end{table*}

Table \ref{table:covcorr} shows correlations between supervised coverage and the \cdc coverage.
All the correlations are very high, with values above $0.9$ on both datasets.
The \hl{correlation scores} are comparable regardless \hl{whether} the supervised coverage uses features based on cosine distance.
This shows that the correlations are not artificially high because the cosine distance is used by both the \sups and \cdc measures.
Interestingly, Pearson correlations are also high, showing that there exists a strong linear \hl{dependency} between the \cdc and the supervised coverage. 
We take the above correlations as evidence that the \cdc measure based on cosine distance 
is indeed a very good unsupervised approximation of supervised topic coverage.
Specifically, high rank correlations show that the \cdc measure can be used to rank topic models by coverage and select the best models.
We therefore proceed to use the \cdc measure, alongside the supervised measure, for evaluation of topic model coverage.

Finally, we give the precise description of the method used to construct the \cdcurve{} and to calculate the values of \cdc measure. 
Given a set of reference topics and a topic model, first a \cdcurve{} is constructed based on cosine distances 
of topic-word vectors. For a specific distance threshold, a reference topic is considered covered if there exists a 
model topic such that its cosine distance from the reference topic is below the threshold. 
The corresponding coverage of the reference set is simply calculated as a proportion of covered reference topics.
The \cdc curve is approximated by segmenting the range \hl{of} possible cosine distances, the $[0, 1]$ interval, 
into $50$ equidistant subintervals. For distances corresponding to subinterval limits
the derived coverages are calculated, which gives a set of $(\mathit{distanceThreshold}, \mathit{coverage})$ points that serve to approximate the curve.
The final \cdc value is calculated from this approximation by using the trapezoidal rule --
the area under a sub-curve corresponding to a subinterval is approximated 
by the area of the trapezoid defined by the interval boundaries and the corresponding coverage values. 
The final \cdc value for the entire curve is then obtained as the sum of the areas of individual trapezoids.

\hl{
The first step in the calculation of the} \hla{\cdc} \hl{measure is the construction of the 
matrix containing cosine distances between the reference and the model topics. 
The time complexity of this operation is $\mathcal{O}(RTV)$, where $R$ is the number of reference topics, 
$T$ is the number of model topics, and $V$ is the vocabulary size.
Once the distance matrix is constructed, the measure can be computed in 
$\mathcal{O}(RK)$ time, where $K$ is the number of distance subintervals.
Namely, if for each reference topic the distance to the closest model topic is stored, 
then for each distance threshold the coverage can be computed in $\mathcal{O}(R)$ time.
Therefore the asymptotic complexity of computing the} \hla{\cdc} \hl{measure is $\mathcal{O}(RTV+RK)$, 
which is equal to $\mathcal{O}(RTV)$ because $K$ is a small constant value.
}

\section{Coverage-Based Model Evaluation}
\label{sect:modelcoverage}

\begin{figure*}[h]

\centering
\includegraphics[width=2\columnwidth]{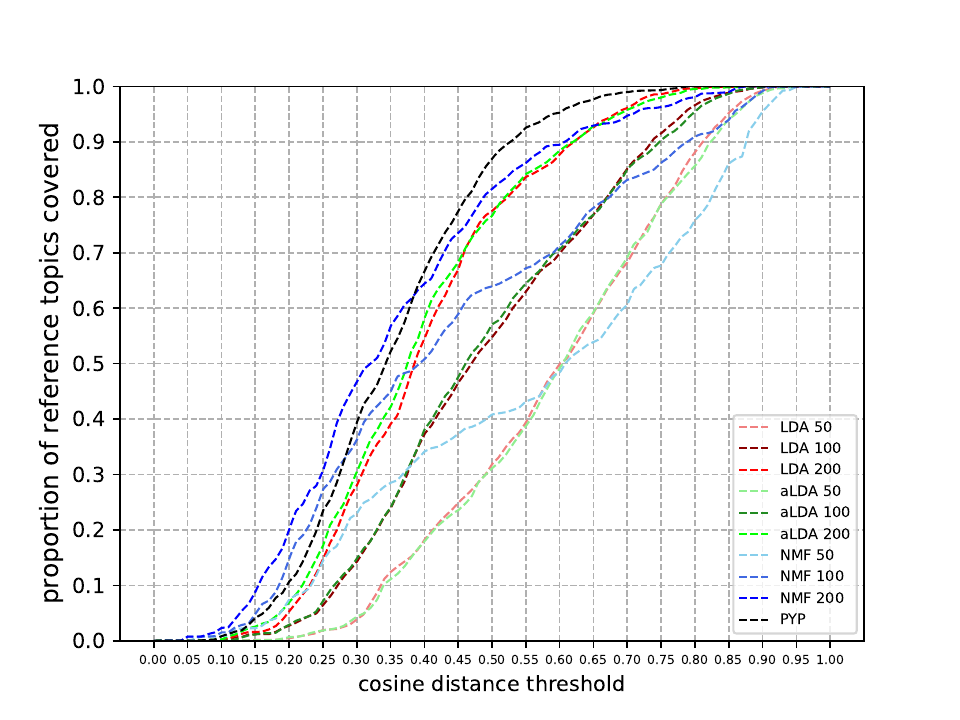}
\caption{\footnotesize Coverage-distance curves depicting coverage of news reference topics by the topic models.} 
\label{fig:ctcalluspol}
\end{figure*}

In this section we apply the proposed measures of topic coverage to analyze the performance of a set of topic models of different sizes and types.
The reference topics are constructed by human inspection and selection of topics learned by topic models,
and represent topics both within the reach of topic models and useful to a human analyst.
Therefore the models are evaluated from the perspective of a use case of topic discovery
on two sets of topics -- topics \hl{occurring} in news articles and biological topics corresponding to phenotypes.

For each dataset four types of topic models are evaluated -- the widely used \ldamodel, its variant \aldamodel, 
the popular \nmfmodel based on matrix factorization, and the nonparametric \pypmodel designed to learn the number of topics.
These models are described in more detail in Section \ref{section:covmodels}.
For each of the parametric models, three different configurations of the 
number of topics are evaluated -- $50$, $100$, and $200$ topics.
These numbers correspond to, respectively, a number smaller than, roughly equal, and larger than the number of \hl{topics} in the reference set.
Ten instances with different random seeds are built per combination of a model type and a number of topics, 
yielding a total of $100$ topic model instances per dataset.

\begin{figure*}[h] 
\centering
\includegraphics[width=2\columnwidth]{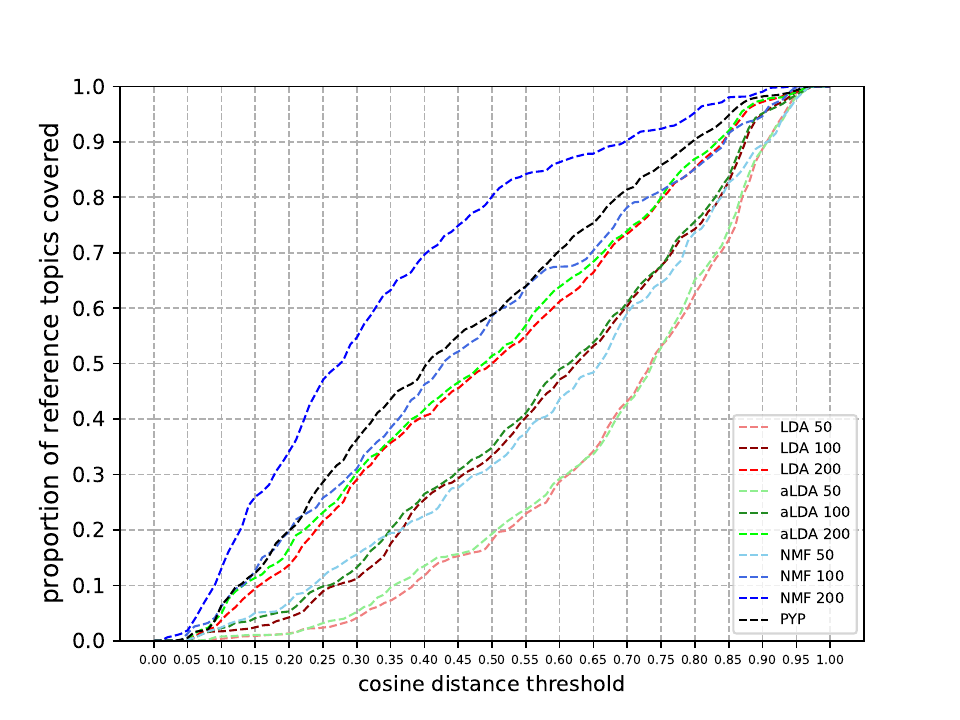}
\caption{\footnotesize Coverage-distance curves depicting coverage of biological reference topics by the topic models.} 
\label{fig:ctcallpheno}
\end{figure*}

Coverage is measured using the \sups measure based on supervised matching of topics.
The result of the supervised coverage is simply the proportion of reference topics covered, i.e., matched by at least one model topic.
The unsupervised \cdc measure is designed to approximate \sups for the purpose of ranking and selection of top models
and is based on the \cdcurve{} which can serve as a standalone graphical tool for analysis and comparison of model coverage.
For each of the measures and for each combination of a model type and a number of topics, 
coverages of the $10$ topic model instances are calculated and averaged to achieve more robust approximations.
The $95\%$ bootstrap confidence intervals of the coverage means
are calculated using the percentile method and $20.000$ bootstrap samples.
Coverage results are shown in the Table \ref{table:coverage} and 
the \cdcurve{s} are shown in Figure \ref{fig:ctcalluspol} and Figure \ref{fig:ctcallpheno}.

Coverage results for different topic models vary depending on the dataset.
On the news dataset, the nonparametric \pypmodel model has the best coverage, 
followed by the \nmfmodel model with $200$ topics.
The \nmfmodel model has the best coverage results among the parametric models,
as shown by both measures and the \cdcurve{s} in Figure \ref{fig:ctcalluspol}
which illustrate how the \nmfmodel models \hl{outperform} the \ldamodel and \aldamodel models with the corresponding number of topics. 
Comparison of the best model \pypmodel and the second best model \nmfmodel-200
shows that while \pypmodel has higher \sups score, the \cdc scores of the two models are the same.
This can be explained by the comparison of the two models' \cdcurve{s} showing 
that the \nmfmodel-200 achieves better coverages for small cosine thresholds, 
while the \pypmodel achieves better coverages for thresholds above $0.37$.
In other words, the \nmfmodel contains more topics that match the reference topics closely, i.e., at smaller cosine distances. 
This means that the \nmfmodel discovers more reference topics at higher level of precision, 
but the superior \sups score of the \pypmodel indicates that its topics 
are still precise enough to be considered as semantically matching. 

On the biological dataset the \nmfmodel model achieves much better coverage results than the probabilistic models.
The \nmfmodel model with $200$ topics has the best overall coverage while for the other model sizes
the \nmfmodel models yield better coverages than the corresponding probabilistic models.
The nonparametric \pypmodel model achieves best results among the low-performing probabilistic models 
and it is comparable with the \nmfmodel models with $100$ topics.
Regardless of the model type, the coverage scores are lower than on the news dataset,
which shows that the biological dataset represents a more challenging scenario of topic coverage. 
On both the news and the biological datasets the coverage score correlates positively with the number of topics
-- larger models are able to uncover more topics.

The coverage results support the claim that the \nmfmodel model is a good default choice for topic discovery.
Namely, on the news dataset the \nmfmodel models outperform the probabilistic \ldamodel and \aldamodel models for all the model sizes, 
while the \nmfmodel model with $200$ topics achieves scores competitive with 
the best-performing nonparametric \pypmodel model configured with the total capacity of $300$ topics.
On the biological dataset the \nmfmodel is clearly the best choice, while the probabilistic models, 
with the exception of the \pypmodel, have weak coverage scores.
These results suggest that the \nmfmodel model is a more robust topic 
discovery tool\hl{, likely} to perform well on different datasets.
The nonparametric \pypmodel model has the best coverage score on the media dataset, 
while on the biological dataset is has the best results among the probabilistic models.
These results support the claim that \pypmodel is a better choice than the nonparametric
\ldamodel and \aldamodel models, \hl{since it is} expected to achieve better coverage and be more robust to dataset change.
Based on the results, the \nmfmodel model with $200$ topics is a good default
choice for performing topic discovery on corpora with between several thousands and several tens of thousands of texts.
The results also show that the \nmfmodel is superior to \ldamodel, regardless of model size.
However\hl{,} in practice the \ldamodel model is very often a first choice, 
probably due to tradition and wide availability of implementations.
The experiments also demonstrate that the unsupervised \cdc measure performs well in
selection of high-performing topic models and demonstrate the use of \cdcurve{} for more in-depth analysis of model coverage.

This section \hl{demonstrates} the merits of coverage-based topic model evaluation 
and demonstrated the application of the proposed coverage methods for model analysis and the selection of high-performing topic models.
However, we note that the proposed methods should be further evaluated 
through their application on additional datasets and topic modeling settings.
We believe that qualitative evaluations focused on human examination of topics would 
reveal \hl{useful} information about both the nature of the model coverage and the measures' performance.
However, such topic evaluations are time consuming and potentially require expert knowledge.
\hl{Appendix} \hla{\ref{app:covapp}} \hl{supplements the experiments in this section
with an analysis of the relationship between topic models' precision and recall,
and with an analysis of the running time of the coverage measures.} 

\newpage

\section{Coverage of Topics Divided into Size Categories}
\label{sect:covbysize}

\begin{table*}[ht]
\caption{\footnotesize Division of reference topics into \hl{quartiles} by size.
\hl{Topic} size is defined as the number of documents containing a topic. 
Each quartile is described by the defining range of documents sizes and by the number of topics it contains.}
\label{table:conceptsizes}
\normalsize
\centering
\begin{tabular}{lllllll}
\toprule
     &            & Q1   & Q2     & Q3      & Q4      & all topics \\
\midrule
News dataset & size & 0--67 & 67--112 & 112--167 & 167--437 &  0--437       \\
     & num. topics  & 34   & 33     & 33      & 33      & 133          \\
\midrule
Biological dataset  & size & 0--9  & 9--20   & 20--42   & 42--216  & 0--216        \\
     & num. topics     & 32   & 26     & 27      & 27      & 112         \\
\bottomrule
\end{tabular}
\end{table*}

The size of the reference topics varies in the sense that some topics occur 
in a large percentage of text documents, while other topics can be found only in a small fraction of documents.
In this section we apply the supervised \sups measure to investigate how topic models cover reference topics of different sizes.
This experiment is partly motivated by several articles in which 
the authors claim that in order to cover smaller topics, 
one needs models configured with a large number of topics \citinl{Wang2015}, 
models that explicitly perform topic diversification \citinl{Xie2015}, or nonparametric models \citinl{Xie2015}.
Additional motivation stems from the observation that in the process of topic discovery
both small and large topics can be of interest to the analyst.
Therefore, \hl{failure} to cover small topics can be a potential drawback of a topic model.

We \hl{define the} size of a reference topic as the number of documents in which the topic occurs,
and that a topic occurs in a document if at least $10\%$ of the document's text is dedicated to the topic.
This heuristical definition is based both on common sense notion of \hl{occurrence} of topics in texts, 
as well as on the basic assumption of probabilistic topic models clearly encoded in the structure of the LDA model \citinl{Blei2003}.
This assumption states that each document is a probabilistic mixture of a set of topics,
and each word in the document ``belongs to'', or talks about, one of these topics.

We proceed to measure the size of reference topics, represented as a weighted lists of words and documents, in the following way.
For each dataset, we use an LDA model that supports both fixed topics and learnable topics.
We build such a model with fixed topics configured to correspond to the reference topics, 
and with additional $20$ learnable topics added for flexibility, i.e., for better approximation of the overall topical structure.
Total number of the model's topics $T$ is therefore $20$ plus the number of reference topics in a dataset.
Document-topic and word-topic distribution of the fixed topics simply correspond to the 
document-topic and word-topic weights of the reference topics normalized to a probability distribution.
Inference is performed by standard Gibbs sampling \citinl{Griffiths2004}, 
with the parameter $\beta$ set to $0.01$ and parameter $\alpha$ set to $1/T$.
Probability formulas used in Gibbs inference are modified in a straightforward way
by insertion of the known topic-word and topic-document weights of the fixed topics.
The learning process of $1000$ Gibbs sampling \hl{iterations} results in learned
probabilities of \hl{occurrence} of fixed reference topics in the corpus documents.
Finally, for each reference topic, the size is calculated as the number of documents
in which the topic occurs with the probability of at least $10\%$.

Reference topics are then divided into quartiles according to the calculated sizes.
This is a principled, data-independent method of division of topics into four categories of approximately equal size. 
The first quartile contains the smallest topics (bottom $25\%$), while the fourth quartile contains the largest topics (top $25\%$).
Table \ref{table:conceptsizes} contains sizes and boundaries of the quartiles for each dataset.
It can be seen that the news dataset with the larger document corpus
has larger reference topics,i.e., news topics tend to occur in more documents than the biological topics.

\begin{table*}[ht]
\caption{\footnotesize Coverage of size categories of reference topics, for both datasets. 
Size categories are defined as \hl{topic size quartiles,} and topic size is \hl{defined as} the number of documents containing the topic. 
For each coverage score, a $95\%$ bootstrap confidence interval is shown.}
\label{table:coveragebysize}
\centering
\normalsize
\begin{tabular}{llllll}
\toprule
     \multicolumn{6}{c}{News dataset}        \\
        & Q1 & Q2 & Q3 & Q4 & All topics \\ 
\midrule
\ldamodel-50   & 0.00 \sgmnt{0.00}{0.00}  & 0.01  \sgmnt{0.00}{0.02}          & 0.14 \sgmnt{0.11}{0.16}            & 0.40  \sgmnt{0.37}{0.42}            & 0.14 \\
\ldamodel-100  & 0.06  \sgmnt{0.05}{0.08}        & 0.14 \sgmnt{0.11}{0.15}            & 0.45 \sgmnt{0.41}{0.50}            & 0.57  \sgmnt{0.55}{0.60}           & 0.31  \\ 
\ldamodel-200  & 0.34 \sgmnt{0.31}{0.37}         & 0.40  \sgmnt{0.39}{0.42}           & 0.64 \sgmnt{0.61}{0.66}            & 0.49  \sgmnt{0.47}{0.51}           & 0.47 \\
\aldamodel-50  & 0.00 \sgmnt{0.00}{0.00}            & 0.01 \sgmnt{0.00}{0.02}           & 0.11 \sgmnt{0.09}{0.14}            & 0.32 \sgmnt{0.29}{0.35}            & 0.12  \\
\aldamodel-100 & 0.08 \sgmnt{0.06}{0.10}         & 0.13 \sgmnt{0.11}{0.15}           & 0.38 \sgmnt{0.33}{0.42}            & 0.45 \sgmnt{0.41}{0.48}            & 0.27  \\
\aldamodel-200 & 0.34 \sgmnt{0.30}{0.38}         & 0.34 \sgmnt{0.32}{0.36}           & 0.60 \sgmnt{0.57}{0.63}             & 0.39 \sgmnt{0.36}{0.42}            & 0.42  \\
\nmfmodel-50   & 0.06 \sgmnt{0.06}{0.06}         & 0.04 \sgmnt{0.03}{0.05}           & 0.34 \sgmnt{0.32}{0.37}            & 0.42 \sgmnt{0.39}{0.44}            & 0.22  \\
\nmfmodel-100  & 0.15 \sgmnt{0.14}{0.17}         & 0.30 \sgmnt{0.27}{0.32}            & 0.65 \sgmnt{0.63}{0.67}            & 0.48 \sgmnt{0.47}{0.49}            & 0.40   \\
\nmfmodel-200  & 0.50 \sgmnt{0.48}{0.52}          & 0.54  \sgmnt{0.52}{0.55}          & 0.73  \sgmnt{0.71}{0.75}           & 0.40 \sgmnt{0.38}{0.41}             & 0.54  \\
\pypmodel     & 0.65 \sgmnt{0.61}{0.68}         & 0.55 \sgmnt{0.52}{0.57}           & 0.79 \sgmnt{0.76}{0.81}            & 0.56 \sgmnt{0.52}{0.59}            & 0.64  \\
\midrule
  \multicolumn{6}{c}{Biological dataset}  \\
  & Q1 & Q2 & Q3 & Q4 & All topics \\
\midrule
\ldamodel-50 & 0.00 \sgmnt{0.00}{0.00}            & 0.00 \sgmnt{0.00}{0.00}            & 0.01 \sgmnt{0.00}{0.03}          & 0.02 \sgmnt{0.01}{0.03}           & 0.01 \small{} \\
\ldamodel-100 & 0.01 \sgmnt{0.00}{0.02}        & 0.03 \sgmnt{0.03}{0.04}         & 0.11 \sgmnt{0.10}{0.12}          & 0.13 \sgmnt{0.09}{0.16}           & 0.07  \small{}\\
\ldamodel-200 & 0.04 \sgmnt{0.02}{0.06}        & 0.14  \sgmnt{0.12}{0.16}        & 0.19 \sgmnt{0.16}{0.22}          & 0.26 \sgmnt{0.22}{0.31}           & 0.16  \small{}\\
\aldamodel-50 & 0.00 \sgmnt{0.00}{0.00}           & 0.00  \sgmnt{0.00}{0.01}           & 0.00 \sgmnt{0.00}{0.00}             & 0.03 \sgmnt{0.02}{0.05}           & 0.01  \small{}\\
\aldamodel-100 & 0.01 \sgmnt{0.00}{0.02}        & 0.05   \sgmnt{0.03}{0.07}       & 0.06 \sgmnt{0.05}{0.07}          & 0.14 \sgmnt{0.12}{0.17}           & 0.07  \small{}\\
\aldamodel-200 & 0.06 \sgmnt{0.05}{0.07}        & 0.14 \sgmnt{0.12}{0.16}         & 0.18 \sgmnt{0.15}{0.21}          & 0.22 \sgmnt{0.20}{0.25}           & 0.15  \small{}\\
\nmfmodel-50  & 0.03  \sgmnt{0.03}{0.03}       & 0.03 \sgmnt{0.03}{0.04}         & 0.10 \sgmnt{0.08}{0.11}           & 0.33 \sgmnt{0.31}{0.35}           & 0.11  \small{}\\
\nmfmodel-100 & 0.04 \sgmnt{0.02}{0.06}        & 0.17  \sgmnt{0.16}{0.19}        & 0.26  \sgmnt{0.25}{0.28}         & 0.40  \sgmnt{0.39}{0.41}           & 0.22  \small{}\\
\nmfmodel-200 & 0.33  \sgmnt{0.30}{0.36}       & 0.52 \sgmnt{0.50}{0.53}         & 0.44 \sgmnt{0.41}{0.45}          & 0.44  \sgmnt{0.43}{0.46}          & 0.44  \small{}\\
\pypmodel & 0.13  \sgmnt{0.11}{0.15}       & 0.21 \sgmnt{0.19}{0.23}         & 0.28 \sgmnt{0.25}{0.31}          & 0.31  \sgmnt{0.29}{0.33}          & 0.23  \small{}\\

\bottomrule
\end{tabular}
\end{table*}

The results showing how the topics in different size quartiles are 
covered by the topic models are displayed in the Table \ref{table:coveragebysize}.
The same topic models as in the coverage experiments in Section \ref{sect:modelcoverage} are used.
The coverages are calculated in the same way as in Section \ref{sect:modelcoverage} --
coverages of $10$ different topic model instances are averaged
and the $95\%$ bootstrap confidence intervals are calculated using the percentile method. 
The results show that the larger models with more topics can cover both
large and small reference topics, while the smaller models can cover only larger topics.
The results for the biological dataset show how the low performance of 
probabilistic models relates to topic size -- these models struggle with covering smaller topics.  
On the other hand, the \nmfmodel achieves much better coverage of smaller topics.
The nonparametric \pypmodel covers smaller topics better than the nonparametric probabilistic models
and has the best coverage over all of the size categories on the news dataset, 
but it lags behind the \nmfmodel on the biological dataset.

The relation of coverage and topic size can be interpreted by looking at the structure of topic models.
Namely, topic models approximate the corpus, represented as the document-word matrix, as a product of the document-topic and topic-word matrices.
Furthermore, these models are learned with the goal of optimizing the reconstruction of the corpus data from the small set of topics. 
We note that while the \nmfmodel model is explicitly based on matrix factorization, 
the described factorization is also in effect performed by the probabilistic topic models \citinl{Roberts2015}.
Therefore, the models with a limited number of topics can achieve better approximation of the text data by 
learning only larger topics that occur in more documents and thus capture more of the data.
On the other hand, large models have additional capacity for fine-grained approximation
and thus can capture both large and small topics.

The results demonstrate that the smaller topic models can successfully cover only the large 
reference topics while the larger models are able to cover both large and small topics.
These results support the previous conjectures that large models and nonparametric topic 
models are needed in order to cover smaller topics \citinl{Wang2015, Xie2015}.
From a practical perspective, these results support the use of larger topic models for topic discovery
since these models can be used both for detection of salient topics and 
for pinpointing small topics which can be of interest to an analyst.
Concretely, in case of the news dataset, the examples of potentially interesting small topics 
from the first size quartile are topics that can be labeled as ``War in Yemen'' and ``Transgender''.
In case of the biological dataset, the reference topics of all sizes represent phenotypes discovered from biological text \citinl{Brbic2016}.
On the other hand, the potential problem with large models is, in our experience, a relatively large number of low quality topics. 
Examples of low quality topics are noisy topics containing random words and documents, and fused topics corresponding to two concepts.
A possible remedy for this problem is the augmentation and speed-up of topic inspection process using measures of topic quality. 
One way to achieve this is to order the model topics by coherence and let the analyst inspect the coherent topics first \citinl{korencic2018}.
Finally, we note that in order to further support the results in this section,
new experiments on other datasets and with other topic model types should be performed.

\section{Coverage of Semantic Categories}
\label{sect:covbytype}

\begin{table*}
\caption{\footnotesize Coverage, by models of varying types and sizes, of news \hl{topics} divided into semantic categories according to two criteria:
topic abstractness and correspondence of a topic to an issue. For each coverage score, a $95\%$ bootstrap confidence interval is shown.}
\label{table:semanticov}
\centering
\normalsize
\begin{tabular}{llllll}
\toprule
        & Abstract & Concrete & Issue & Non-issue & All topics \\
        \cmidrule(lr){2-3} \cmidrule(lr){4-5} \cmidrule(lr){6-6}
\ldamodel-50   & 0.13 \sgmnt{0.11}{0.15}    & 0.15 \sgmnt{0.14}{0.16}     & 0.22 \sgmnt{0.21}{0.24}   & 0.09 \sgmnt{0.08}{0.10}  & 0.14 \\
\ldamodel-100   & 0.26 \sgmnt{0.23}{0.29}     & 0.37 \sgmnt{0.35}{0.39}     & 0.48 \sgmnt{0.44}{0.52}  & 0.21 \sgmnt{0.20}{0.23}  & 0.31 \\
\ldamodel-200  & 0.32 \sgmnt{0.30}{0.34}     & 0.62 \sgmnt{0.59}{0.64}     & 0.61 \sgmnt{0.58}{0.64}  & 0.39 \sgmnt{0.37}{0.40} & 0.47 \\
\aldamodel-50  & 0.10 \sgmnt{0.09}{0.12}      & 0.13 \sgmnt{0.11}{0.14}     & 0.19 \sgmnt{0.17}{0.21}  & 0.07 \sgmnt{0.06}{0.08} & 0.12 \\
\aldamodel-100 & 0.20 \sgmnt{0.18}{0.23}      & 0.32  \sgmnt{0.30}{0.35}    & 0.45 \sgmnt{0.42}{0.47}  & 0.15 \sgmnt{0.14}{0.17} & 0.27 \\
\aldamodel-200 & 0.30 \sgmnt{0.28}{0.32}      & 0.54 \sgmnt{0.52}{0.57}     & 0.56 \sgmnt{0.54}{0.59}  & 0.33 \sgmnt{0.30}{0.36} & 0.42 \\
\nmfmodel-50  & 0.13 \sgmnt{0.11}{0.14}     & 0.32  \sgmnt{0.31}{0.33}    & 0.32 \sgmnt{0.31}{0.33}  & 0.17 \sgmnt{0.16}{0.17}  & 0.22 \\
\nmfmodel-100  & 0.26 \sgmnt{0.25}{0.27}     & 0.54 \sgmnt{0.53}{0.55}     & 0.53 \sgmnt{0.52}{0.55}  & 0.32 \sgmnt{0.31}{0.33} & 0.40 \\
\nmfmodel-200  & 0.37 \sgmnt{0.35}{0.38}     & 0.71 \sgmnt{0.70}{0.72}     & 0.63 \sgmnt{0.61}{0.64}  & 0.49 \sgmnt{0.48}{0.50} & 0.54 \\
\pypmodel   & 0.47 \sgmnt{0.44}{0.49} & 0.80 \sgmnt{0.77}{0.83}      & 0.78 \sgmnt{0.75}{0.80}  & 0.55 \sgmnt{0.53}{0.57} & 0.64 \\
\bottomrule
\end{tabular}
\end{table*}

Topic models are useful tools for text exploration and topic discovery since they are 
able to learn topics that humans can interpret as concepts.
When topic models \hl{are} applied in computational social sciences 
it is often \hl{desirable} that model topics correspond to concepts from a specific category.
Such topics of interest to the researcher have been described as 
``theoretically interesting'' \citinl{grimmer2013text} and ``analytically useful'' \citinl{dimaggio2013}.

An example of a research topic which can benefit from quantitative 
analysis of news text based on topic models is agenda setting \citinl{Bonilla2013Dec, kim2014computational, Papadouka2016}. 
Agenda setting research \citinl{mccombs1972agenda} is focused on salience of issues -- topics of political or social importance. 
The standard approach is to investigate how media salience of issues relates to public perception of their importance \citinl{mccombs1972agenda}.
Naturally, when topic models are applied for agenda setting research it is \hl{desirable} that the model topics correspond to issues.
Computational agenda setting studies typically rely on topic models to automatically
detect issues in a collection of news texts and to measure their salience \citinl{Bonilla2013Dec, kim2014computational, Papadouka2016}.
In this section we demonstrate the application of coverage methods 
\hl{to the analysis of} how topic models cover the issues \hl{occurring} in news texts.
Such analyses could guide the choice of topic modelling tools for agenda setting studies.

There exist numerous other research directions, each with its own class of ``\hl{theoretically} interesting'' concepts.
Examples of such studies include the analysis of news framing \citinl{dimaggio2013, Gilardi2018},
analysis of historical news \citinl{Yang2011}, and qualitative analysis of news focused on a specific topic \citinl{Evans2014}.
For these and numerous other use cases an experiment focused on the coverage of topics of interest could be conducted.
Motivated by applications in social sciences where the topics of interest are expectedly abstract concepts, 
we analyze how the \hl{topic} models cover abstract reference topics. 

We proceed to measure the coverage of reference topics divided according to two
criteria -- correspondence to a news issue and topic abstractness.
To this end, each reference topic from the news dataset was annotated 
as being either abstract or concrete, and as being either an issue or 
a non-issue\footnote{We slightly abuse the language semantics and use the term ``non-issue topic'' to denote the 
topic that does not correspond to a news issue, not a topic of little or no importance.} topic.
A topic was considered abstract if it could be interpreted as an abstract concept, 
and it was considered concrete if it could be interpreted as either a person, a country, an organization, or an event.
A topic was defined as corresponding to an issue if it was strongly related to an important social or political issue.
Table \ref{topic-examples} contains interpretable model topics representative of the reference topics. 
The ``Climate Change'' topic is an example of a topic that is both abstract and an issue,
while the ``China'' and ``Boston Bombing Trial'' topics are examples of concrete topics.
Reference topics were annotated by two annotators. 
First\hl{,} a sample of $30$ topics was annotated by both annotators 
and Krippendorph's $\alpha$ \hl{coefficients} of inter-annotator agreement were calculated.
For topics abstractness $\alpha$ was $0.67$, and for issue vs. non-issue labels $\alpha$ was $0.44$.
The levels of agreement reflect the fact that the assessment weather topic corresponds to an issue 
is more difficult and open to interpretation than the relatively straightforward \hl{assessment} of topic abstractness.
In the next step the topics for which annotators' assessments differred were discussed, 
which led to an improved understanding of the definitions guiding the annotation process.
Annotations of the $30$ sampled topics were synchronized, after which each annotator 
proceeded to annotate half of the remaining topics -- approximately $50$ topics per annotator. 
After the annotation was completed, each of the $133$ reference topics from the news dataset 
was labeled both as being either abstract or concrete, and as corresponding to a news issue or not.

The coverage of the resulting topic categories by topic models was then measured. 
Topic models used in the experiments are the same models used in coverage experiments in Sections \ref{sect:modelcoverage} and \ref{sect:covbysize}.
The coverage was measured using the supervised coverage measure \sups based on the requirement of a precise match between the model and reference topics.
The coverages are calculated in the same way as in Section \ref{sect:modelcoverage} --
coverages of $10$ different topic model instances are averaged
and the $95\%$ bootstrap confidence intervals are calculated using the percentile method. 
The coverage results, displayed in Table \ref{table:semanticov}, 
show that the \nmfmodel model has better coverage of the issue topics than the \ldamodel model,
although this advantage becomes smaller as the number of model topics increases.
As for the non-issue topics, the \nmfmodel model is clearly better than the \ldamodel model.
The \pypmodel model achieves the best performance and covers almost $80\%$ of all the issues.
Interestingly, issue topics are covered better than non-issue topics across all model types and sizes. 

The \nmfmodel model is slightly better than the \ldamodel model in coverage of the abstract topics, 
and clearly better in coverage of the concrete concepts.
The nonparametric \pypmodel performs best for both concrete and abstract concepts.
We also observe that the concrete topics are covered better than the abstract topics across all model types and sizes. 
A possible explanation is that the concrete topics focused on people, 
events, and organizations, are more prevalent in the news text and thus more easier to detect. 
Similarly, the higher coverage of the issue topics might be explained by 
the fact that news articles favor issue topics.

The coverage experiment described in this section is motivated by applications of topic models in social sciences, 
where models are expected to correspond to concepts of interest to the researcher. 
In such studies \ldamodel is often a model of choice, due to tradition and availability of \ldamodel implementations.
However, experiments in this section suggest that the \nmfmodel and \pypmodel models are a better choice, at least in the case of news analysis,
since both models achieve better coverage of abstract topics and issue topics than \ldamodel model.
These results are in line with previous experiments in Sections \ref{sect:modelcoverage} and \ref{sect:covbysize}
showing that \nmfmodel and \pypmodel models achieve higher overall coverage and are
better at pinpointing smaller topics than the \ldamodel model.
While the nonparametric \pypmodel model achieves the best overall coverage of all the topic categories, 
in our opinion the \nmfmodel model is a better choice.
We base this assessment on evidence in Section \ref{sect:modelcoverage} 
that shows greater robustness of \nmfmodel performance \hl{across} datasets.
\hl{Additionally}, regardless of model type, models with more topics are a better choice, 
probably due to their ability to detect small topics, as suggested by the results in Section \ref{sect:covbysize}.

The experiment in this section is a demonstration how the coverage-oriented model evaluation
can be applied to analyze and select topic models best suited for the 
purpose od topic discovery in social sciences.
Such analyses rely on measures of topic coverage and use-case oriented 
sets of reference topics that represent concepts of \hl{interest}.
However, to obtain reliable and generalizable results, similar experiments should be \hl{performed}
for more \hl{use cases} representing different research designs.
Ideally, such experiments would generate enough evidence for reliable recommendations for use of specific topic model types.
The findings in this section also show variations in coverage of different semantic topic categories.
We find a more in-depth investigation of this phenomenon an interesting topic for future work 
with potential to generate knowledge about the structure of conceptual topics and models expected to approximate them.

\section{Coverage and other Topic Model Evaluation Methods}
\label{sect:covcohstabil}

In this section we examine how the proposed coverage approach 
relates to two other topic model evaluation methods -- \emph{topic coherence} and \emph{topic model stability}. 
The experiments demonstrate that the topic coverage is a property distinct from both coherence and stability.
In the case of model stability, we show how the coverage measures can be adapted to approximate model stability.

\subsection{Coverage and Topic Coherence}
\label{sect:covcoh}

Topic coherence \citinl{Newman2010} is an approach to evaluation of topic models
based on calculating a measure of coherence of individual topics. 
A good coherence measure is correlated with topic interpretability
in the sense of the topic's correspondence to a single concept \citinl{Newman2010}.
Topic coherence measures typically use top-weighted topic words as input,
compute coherence by aggregating mutual similarity of top words,
and are designed to maximize correlation with human coherence scores.
Calculation of topic coherence became a popular method of topic model evaluation 
and many coherence measures have been proposed \citinl{Roeder2015}. 

Coverage is related to coherence because both approaches aim to approximate the matching between model topics and concepts.
However, coherence is more generally and loosely defined as a measure of match between a topic and any concept, 
\hl{while coverage} is defined in terms of a predefined set of specific concepts represented by reference topics.
Coherence measures rely only on topic-related words and a model of word similarity and are thus easier to 
deploy than coverage measures that \hl{require a} set of pre-compiled reference topics. 
Therefore, coherence measures are more approximative but readily available measures of topic conceptuality, 
while coverage provides more precise evaluation at the added cost of effort needed to construct reference topics.
In this section we experimentally examine the relation between the two approaches 
by calculating correlations between coherence measures and measures of coverage proposed in Section \ref{section:measures}.

Coherence measures calculate coherence scores of model topics 
and in order to enable comparison of coherence and coverage measures, 
we adapt the coverage measures to compute coverage-related scores of individual topics. 
The adaptation is performed by using the existing topic-matching criterion of a coverage measure to compare a model topic with reference topics.
The adapted measures thus score individual model topics in terms of their correspondence with the reference topics, 
which is a straightforward application of existing coverage apparatus to topic-level scoring.

\begin{table*}[h]
\caption{\footnotesize Spearman rank correlations between coherence and coverage measures, 
calculated on 13.500 topics for each of the datasets. For each correlation \hl{coefficient}, a $95\%$ bootstrap confidence interval is shown.}
\centering
\normalsize
\setlength{\tabcolsep}{4pt}
\begin{tabular}{lrlrlrlrl}
\toprule
 & \multicolumn{4}{c}{News dataset}              &         \multicolumn{4}{c}{Biological dataset}         \\
 \cmidrule(lr){2-5} \cmidrule(lr){6-9}
 Coherence      &  \multicolumn{2}{c}{\sups}      & \multicolumn{2}{c}{\cdc} & \multicolumn{2}{c}{\sups} & \multicolumn{2}{c}{\cdc} \\
\midrule
\npmil{wiki}  & -0.06 & \sgmnt{-0.08}{-0.05}      & 0.02 &  \sgmnt{0.00}{0.04}    & -0.05 & \sgmnt{-0.07}{-0.04} & 0.18 & \sgmnt{0.16}{0.19}  \\
\npmil{corpus} & 0.29 & \sgmnt{0.28}{0.31}    & 0.48 & \sgmnt{0.46}{0.49}  & -0.05 & \sgmnt{-0.07}{-0.03}    & 0.10 & \sgmnt{0.08}{0.12}   \\
\cpl{wiki}   & -0.03 & \sgmnt{-0.05}{-0.01}    & 0.10 & \sgmnt{0.08}{0.12}  &    -0.16 & \sgmnt{-0.18}{-0.15}    & 0.00 &  \sgmnt{-0.02}{0.02}    \\
\cpl{corpus}   & 0.35 &  \sgmnt{0.33}{0.36}      & 0.50 & \sgmnt{0.48}{0.51}     &   -0.02 & \sgmnt{-0.04}{0.00}    & 0.19 & \sgmnt{0.17}{0.21}  \\
\cvl{wiki}    & 0.14  & \sgmnt{0.13}{0.16}      & 0.22  & \sgmnt{0.20}{0.24}   &  -0.16 &  \sgmnt{-0.18}{-0.14}   & -0.10 & \sgmnt{-0.12}{-0.08}   \\
\cvl{corpus}   & 0.23  & \sgmnt{0.22}{0.25}      & 0.19 &  \sgmnt{0.18}{0.21}   &   0.27 &  \sgmnt{0.25}{0.28}    & 0.20 & \sgmnt{0.19}{0.22}  \\
\bottomrule
\end{tabular}
\label{table:covcohcorr}
\end{table*}

Concretely, in case of the supervised coverage\hl{,} the score of a model topic
is set to $1$ if the topic matches a reference topic or to $0$ if no matching reference topic exists.
The matches are computed using the supervised topic matcher described in Section \ref{section:supmeasures}.
The unsupervised \cdc measure, described in Section \ref{section:cdc}, is based on approximation 
of equality between a model topic and a reference topic using a cosine distance threshold -- 
the topics are considered equal if their cosine distance is below the threshold.
We adapt the \cdc for calculating the match between a single model topic 
and the reference topic set by using cosine similarity, the inverse of cosine distance.
Specifically, we compute the cosine similarity between a model topic and the set of reference topics, 
i.e., the similarity between the topic and its most similar reference topic.  

\subsubsection{Measures of Topic Coherence}

We compare the coverage measures with state of the art coherence measures 
that achieved top correlations with human coherence scores in an extensive evaluation experiment \citinl{Roeder2015}. 
In \citinl{Roeder2015}, a generic structure of a coherence measure is formulated
and the derived space of possible measures is searched for top performing candidates.
The topic coherence scores are calculated by dividing the set of top topic words 
into subsets and aggregating the similarities between word subsets. 
Within this framework, the averaging of similarities between individual words is a special case.
A coherence measure also depends on a model of word similarity derived from 
word \hl{co-occurrence} counts calculated using either the local corpus or the Wikipedia.

We evaluate three of top-performing measures -- newly discovered measures labeled \cp and \cv \citinl{Roeder2015}, 
and the previously proposed \npmi measure of \citnoun{Aletras2013}. 
The \npmi measure is calculated by averaging normalized pointwise mutual information similarity of word pairs.
The \cp measure averages similarities, defined using conditional probability \citinl{Fitelson2003}, 
between pairs consisting of a word and its complement set.
The \cv measure averages similarities between all pairs consisting of a top topic word and
the set of those top words with higher topic-word \hl{weights}.
In case of \cv, similarities are \hl{calculated} by representing words as vectors of similarity scores with other top topic words.

For each of the described measures we experiment with both approaches to defining word similarity --
domain-specific similarity derived from word \hl{co-occurrences} in the local corpus,
and the generic mixed-domain similarity based on \hl{co-occurrences} in the English Wikipedia.
All the coherence measures in this experiment use as input top $10$ topic words \citinl{Roeder2015}.

\subsubsection{Correlation Between Coverage and Coherence}

We proceed to \hl{experimentally} evaluate the relationship between the described coherence 
and coverage measures by calculating the Spearman rank correlations.
Correlations are calculated on the set of topics of topic models described in Section \ref{section:covmodels}.
These models of different sizes and types yield a total of $13.500$ topics per dataset.
The $95\%$ bootstrap confidence intervals of the Spearman correlation coefficients 
are calculated using the percentile method and $20.000$ bootstrap samples.
The results, presented in Table \ref{table:covcohcorr}, show that neither strong 
nor consistent correlation between coherence and coverage exists.
The correlation strength varies, depending on the measure type and dataset, and in most cases it is weak to non-existent.

The results in Table \ref{table:covcohcorr} show that in majority of cases coverage correlates better
with corpus-based coherence measures than with Wikipedia-based coherence measures.
A possible explanation is that reference topics are derived from the topics of models built on corpus texts.
Namely, probabilistic topic models implicitly assume that words defining a topic 
are the words that tend to co-occur in texts \citinl{dimaggio2013}.
Therefore topics with high local coherence, being by definition topics whose top-words co-occur in corpus texts,
should have better likelihood of being among the reference topics.

We also observe that the \cdc measure correlates better with coherence than the \sups measure based on supervised topic matching.
A reasonable explanation is that \cdc is by design less error-prone to correlation errors 
since it calculates an approximative similarity score between a model topic and reference topics.
Namely, if a coherent topic is missing from the reference set, its cosine similarity 
to the set can still be high if other similar reference topics exist. 
And if the supervised matcher does not recognize a coherent topic as matching with 
a corresponding reference topic, cosine similarity of these two topics is still expected to be high.

Finally, we observe that the correlations between coherence and coverage 
are higher in case of the News dataset, while for the Biological dataset they are very weak at best.
A reasonable explanation is that the News reference topics are more representative of the set of all the learnable topics. 
Namely, both sets of reference topics are derived from model topics but
Biological topics are filtered to include only topics describing phenotypes.
Therefore, non-phenotype topics with high coherence scores 
are likely to receive low coverage scores that can negatively affect the correlation strength.

\subsubsection{Conclusions}

The experiments in this section show that the coverage and coherence clearly differ,
although both approaches measure a correspondence between model topics and concepts.
In particular the proposed coverage measures, designed to evaluate topic models in terms
of their match with reference topics based on concrete topic discovery use cases, 
cannot be approximated well with state-of-the-art coherence measures.

These findings open interesting questions about the relation between coherence and coverage measures.
We believe that an in-depth investigation of relationships between coverage and coherence could improve both evaluation approaches.
Concretely, it would be interesting to investigate the correlations between the two 
and observe which coherence measures best correlate with coverage of which sets of reference topics.
This might better explain the nature of the vaguely defined property of topic coherence 
by comparing it with precisely defined coverage measures.
\hl{Additionally}, the coherence measures able to approximate coverage well could be used for \hl{selection of interpretable models}.

Our findings are in line with the previous work on the analysis of coherence measures.
Experiments in \citinl{Chuang2013} show weak and inconsistent correlation between 
measures of topic quality and a measure of correspondence between topics and reference concepts,
with the coherence measures exhibiting mildly negative correlations.
Evaluation of coherence measures via topic-level quality scores based on human
interpretation of topics was performed in \citinl{doogan2021}.
The experiment found neither consistent nor significant correlation between 
coherence measures on one, and the interpretability scores based on 
inspection of topic words and documents on the other side.
Both our experiment and the two previous experiments 
show variations in performance of different types of coherence measures.
All the experiments also show weak correlation of coherence with interpretable 
measures of quality grounded in either reference topics or human labels.
Both the correlations in Table \ref{table:covcohcorr} and those in \citinl{doogan2021} tend to be weakly or \hl{moderately} positive. 
This indicates that coherence measures can roughly approximate topic interpretability, 
but fail to calculate scores that are reliable proxies for interpretability.

\subsection{Coverage and Model Stability}
\label{sect:covstabil}

In this section we examine the relationship between topic coverage and model stability.
In the first experiment we measure correlation between the measures of stability and coverage. 
Next, we show how the proposed coverage measures can be adapted to calculate stability.
This adaptation relies on the fact that both approaches are based on approximation of a match between two topics.

\begin{table*}[h]
\caption{\footnotesize Spearman correlations between the coverage and stability measures, 
as well as between \hl{each of the measures and the number of model topics}.
Correlations are calculated on $80$ sets of $10$ model instances. For each correlation \hl{coefficient}, a $95\%$ bootstrap confidence interval is shown.}
\centering
\normalsize
\setlength{\tabcolsep}{3pt}
\begin{tabular}{lrlrlrlrl}
\toprule
          & \multicolumn{4}{c}{News dataset}  & \multicolumn{4}{c}{Biological dataset}         \\
\cmidrule(lr){2-5} \cmidrule(lr){6-9}          
           & \multicolumn{2}{c}{\bipstab} & \multicolumn{2}{c}{num. topics} & \multicolumn{2}{c}{\bipstab} & \multicolumn{2}{c}{num. topics} \\
\midrule
\sups      & -0.09 & \sgmnt{-0.32}{0.14}  &  0.57 & \sgmnt{0.34}{0.74}  & 0.10 & \sgmnt{-0.17}{0.33}  &  0.74 & \sgmnt{0.59}{0.85} \\
\cdc       & -0.26 & \sgmnt{-0.50}{-0.00} &  0.79 & \sgmnt{0.63}{0.89} & 0.01 & \sgmnt{-0.27}{0.25} &  0.79 & \sgmnt{0.66}{0.89} \\
\bipstab   &  &  & -0.54 & \sgmnt{-0.71}{-0.33} &  &  & -0.53 & \sgmnt{-0.71}{-0.33} \\      
\bottomrule
\end{tabular}
\label{table:covstabcorr}
\end{table*}

\subsubsection{Topic Model Stability}

A popular approach to the evaluation of topic models is 
based on the notion of model stability \citinl{steyvers2007stabil, DeWaal2008, Greene2014, Koltcov2014, Chuang2015, Belford2018}.
The approach is motivated by the fact that the learned model instances vary randomly
due to model inference algorithms that rely on random initialization and sampling. 
Instability of topic models has been confirmed both numerically \citinl{DeWaal2008, Greene2014, Belford2018} 
and by model inspection that reveals topic variation among model instances \citinl{Koltcov2014, Chuang2015}.
This variation is potentially detrimental for an analyst performing topic discovery \citinl{Belford2018}, 
especially in the case of social sciences where omitted topics and topic variations 
can influence the results of a study \citinl{Koltcov2014, Chuang2015}.
Measures of stability are derived from mutual similarity of inferred model instances 
-- a stable topic modeling setting should consistently produce similar models.
Similarity between models is calculated either by aligning model topics 
using a similarity measure \citinl{DeWaal2008, Greene2014, Koltcov2014, Belford2018}
or by comparing models represented in terms of words or documents \citinl{Belford2018}.
Alternatively, an interactive approach based on clustering and visualizing topics of many models has been proposed \citinl{Chuang2015}.

In the following experiments, we opt for a common approach to stability calculation based 
on aligning topics of two model instances by using a measure of topic similarity \citinl{DeWaal2008, Greene2014, Belford2018}.
The first step of the approach is to find an optimal bipartite matching of topics, 
i.e., an optimal one-to-one pairing between first model's and second model's topics. 
This optimal pairing that maximizes pairwise topic similarity is computed using the Hungarian algorithm \citinl{kuhn1955}.
The similarity of two models is then computed as the average similarity of the paired topics.
Finally, given a topic model and an inference algorithm, the stability 
is calculated as the average mutual similarity of the inferred model instances.
We use the cosine similarity of topic-word vectors as the measure of topic similarity
and we label the described measure of stability as \emph{\bipstab}.

\subsubsection{Correlation Between Coverage and Stability}

The correlation between stability and coverage measures is computed at the level of a set of model instances.
Since the set of topic models described in Section \ref{section:covmodels} contains $10$ instance sets,
we extend it in order to obtain more robust correlation results.
Additional models are built using the same procedure, described in Section \ref{section:covmodels}.
In the extended model set the model size, defined by the number of topics parameter $T$, is varied over a wider range of values. 
For parametric topic models \ldamodel, \aldamodel, and \nmfmodel, 
the parameter $T$ is varied between the values of $20$ and $500$ in steps of $20$, yielding $25$ size variants per model type.
For the nonparametric \pypmodel model, the maximum learnable number of topics $T$ 
is varied between the values of $100$ and $500$ in steps of $100$, yielding five size variants.
The final extended model set contains, for each of the two datasets, $80$ sets of $10$ model instances. 
Each instance set represents a specific model type and size, and contains instances built using different random seeds. 

\begin{table*}[h]
\caption{\footnotesize Spearman correlation between the coverage-based stability measures and the \bipstab measure. 
For each correlation \hl{coefficient}, a $95\%$ bootstrap confidence interval is shown.}
\centering
\normalsize
\begin{tabular}{llllll}
\toprule
  & News dataset & Biological dataset \\
\midrule
\css   & 0.69 \sgmnt{0.51}{0.83}  & 0.69 \sgmnt{0.48}{0.84}  \\         
\cdcs & 0.99 \sgmnt{0.98}{1.00} & 1.00 \sgmnt{0.99}{1.00} \\
\bottomrule 
\end{tabular}
\label{table:stabilcorr}
\end{table*}

We proceed to empirically determine the nature of relation between coverage and stability, 
by calculating Spearman rank correlations between the described \bipstab stability measure
and the two proposed measures of coverage, \sups and \cdc.
The $95\%$ bootstrap confidence intervals of the correlation coefficients 
are calculated using the percentile method and $20.000$ bootstrap samples.
Results, presented in Table \ref{table:covstabcorr}, show that there is no strong correlation between the stability and coverage measures. 
The correlations are weak to non-existent, ranging from slightly negative to slightly positive. 
These results show that model coverage is a property of topic models unrelated to model stability.

The lack of correlation can be explained, at least in part, by the nature of 
correlation between the number of model topics on one, and the stability and coverage on the other side.
As can be seen from Table \ref{table:covstabcorr}, stability has a negative while 
coverage has a positive correlation with the number of topics.
This indicates that the larger models tend to be less stable, which is unsurprising
since they contain more learnable variables, resulting in more variation among the learned model instances.
On the other hand, larger models with the capacity to learn more topics tend to have greater coverage, 
which is in line with the results of the coverage experiments in Section \ref{sect:modelcoverage}.

\subsubsection{Calculating Stability using Coverage Measures}
\label{sect:cov4stabil}

We proceed to show how the coverage measures can be adapted to measure model stability 
by way of direct comparison of two model instances.

First, we adapt the supervised coverage measure \sups based on supervised matching of model and reference topics.
The stability measure derived from \sups calculates the similarity between 
two topic models in terms of the reference topics covered by both models. 
Concretely, given a set of reference topics and a topic matcher, 
let $\mathit{reftop(m)}$ be the set of reference topics that the topics of the model $m$ cover. 
The similarity of two models $m_1$ and $m_2$ with the same number of topics $T$ is defined as $|\mathit{reftop(m_1)} \cap \mathit{reftop(m_2)}|/T$.
In other words, the similarity of two models is defined as the number of reference topics 
discovered by both instances, relative to the maximum number of discoverable topics.

This definition of similarity is related to the standard model similarity based on bipartite matching,
computed as the average similarity between the pairs of highly similar topics of two models. 
Namely, in our case the pairs of similar topics correspond to reference topics discovered by both models
-- each such reference topic matches a topic from the first and a topic from the second model.
Therefore, the number of mutually discovered reference topics can be interpreted as the sum of binary similarities of aligned model topics.

The final stability measure based on \sups is calculated, for a set of model instances,
as the average similarity between instances.
We dub the described measure of stability \emph{reference set stability} and label it as \emph{\css}.

The \cdc measure of coverage can \hl{also} be adapted for measuring model stability
by using it to compute the similarity between two model instances. 
The \cdc measure, defined in Section \ref{section:cdc}, is designed to approximate how well a set of 
topics of a model $m$ covers a set of reference topics $\mathit{ref}$. We denote the corresponding coverage score as $\mathit{\cdcm(ref, m)}$.
The \cdc-based similarity of two topic models $m_1$ and $m_2$ is computed as $(\mathit{\cdcm(m_1, m_2)} + \mathit{\cdcm(m_2, m_1)})/2$. 
In other words, this is an approximation of how well the sets of topics of the two models cover each other.
The coverage is computed in both directions and averaged in order to make the similarity measure symmetrical.
The final stability measure is calculated, as in the case of other stability measures,
by averaging pairwise similarity on a set of model instances.
We dub this measure of stability \emph{\cdc stability} and label it as \emph{\cdcs}.

In order to measure how the coverage-based measures of stability relate to the 
\bipstab measure, we calculate Spearman correlations on the 
extended dataset of models of varying types and sizes divided into $80$ sets of model instances.
The $95\%$ bootstrap confidence intervals of the correlation coefficients 
are calculated using the percentile method and $20.000$ bootstrap samples.

The results from Table \ref{table:stabilcorr} show that 
the \css stability achieves substantial correlation with standard stability, 
while in the case of \cdcs stability the correlation is almost perfect.
This difference is not surprising since the \css measure matches two models via reference topics,
while the \cdcs measure compares the models directly, which leads to a better approximation of similarity.
Namely, in case of the \css measure, if two models contain an identical topic that is not in the reference set,
the similarity will be negatively affected. 
Such a model topic not in the reference set could be either a missing conceptual topic or a commonly \hl{occurring} stopwords or noisy topic.
On the other hand, the \cdcs measure will \hl{successfully} match the same topic \hl{occurring} in models being compared. 

Stability of topic models is, by definition, the property of the modeling setup to consistently produce the same topics.
Since the topic models are expected to produce topics that can be interpreted as concepts, 
this implies that stability can be viewed as a property of models to consistently uncover the same concepts.
The substantial level of correlation between the \css measure and 
the standard \bipstab measure can be interpreted as an experimental confirmation of the previous intuitive claim.
Namely, the \css measure approximates stability with the amount of reference topics, 
corresponding to concepts, uncovered by each of the two distinct model instances.

The correlation between stability based on \cdc and \bipstab is almost perfect. 
The very high level of correlation is a \hl{useful} experimental finding, which we take 
as proof-of-concept for the application of the \cdc measure for calculation of model stability.
This finding has practical benefits, since the \cdcs stability measure 
is much faster to compute than the \bipstab stability measure.

Namely, the \bipstab measure calculates model similarity by optimally aligning the model topics
using the Hungarian algorithm \citinl{kuhn1955} with the computational complexity of $O(T^3)$, where $T$ is the number of model topics. 
On the other hand, the \cdcs measure calculates model similarity using the \cdc measure, described in detail in Section \ref{section:cdc}.
The \cdc measure integrates best-case matching results over a range of distance thresholds
and can be computed with time complexity of $O(T^2)$, which corresponds to the time necessary 
to calculate distances between all pairs of topics.

In practice, the calculation of \cdcs is orders of magnitude faster.
For each of the two datasets, calculation of the \bipstab measure on $80$ sets of model instances took approximately two weeks. 
On the other hand, calculations of the \cdcs measure were completed in under two hours.
Therefore the proposed \cdcs stability measure has a potential to greatly 
speed up stability-based model evaluation and to make such evaluations 
viable for large model collections and models with a large number of topics. 

\subsubsection{Conclusions}

The experiments in this section show that model stability is a property 
of topic models which is unrelated to the model's coverage of a set of reference topics.
In other words, stable models that consistently uncover the same topics
do not necessarily uncover all the useful topics or topics of particular interest to an analyst.
This implies that optimizing topic models using the stability as 
the only criterion might not lead to best quality models.
On the other hand, it is reasonable to expect that a topic model able 
to uncover the majority of topics within its reach would be stable.
We believe that follow-up experiments, on new corpora and news sets of reference topics,
are needed to further investigate the relationship between the stability and coverage.

We also show how the proposed coverage measures can be adapted to calculate stability. 
The adapted measures achieve a good correlation with the standard stability measure,
support the interpretation of stability in terms of consistent uncovering of concepts, 
and provide a computationally efficient alternative to stability calculation.
We note that, while the \cdcs and \bipstab measures use different algorithms for model similarity,
both rely on cosine-based similarity of topic-word vectors for matching of individual topics.
Therefore, in order to fully generalize the approach, experiments with the \cdcs measure variants
based on other measures of topic similarity should be investigated.
Such investigations would also ideally include new datasets and more topic model types.

\section{Related Work} 

\subsection{Topic Models}
Topic models \citinl{Blei2003} are unsupervised models of text capable of learning topics from large text collections. 
Each topic is a construct typically characterized by weighted lists of words and documents
and expected to correspond to a concept occurring in texts.
Topic models have numerous applications, including exploratory text analysis \citinl{Chuang2012}, 
information retrieval \citinl{Wei2006}, feature extraction \citinl{Chen2011},
natural language processing \citinl{boyd2007topic, lin2009joint}, 
and applications in computational social sciences 
\citinl{Bonilla2013Dec, Evans2014, Jacobi2016, Gilardi2018}.

Two prominent families of topic models are probabilistic models \citinl{Blei2012},
such as Latent Dirichlet Allocation \citinl{Blei2003}, and matrix factorization models, such as Nonnegative Matrix Factorization \citinl{Arora2012}.
Generative probabilistic models are a dominant approach to topic modelling. 
These models are based on a probabilistic process of text generation and
their structure is defined in terms of a set of random variables and relations between them.
There exist a variety of probabilistic model types with structure defined by 
random variables corresponding to various text metadata \citinl{Blei2012}.
Unlike the models that assume a fixed number of topics, models relying on Bayesian 
nonparametric inference are able to infer the number of topics from data \citinl{Teh2006, Buntine2014}.

Models based on matrix factorization, such as latent semantic analysis \citinl{Deerwester1990} 
and non-negative matrix factorization \citinl{Lee1999} are a popular alternative to generative models.
These models learn a set of latent factors, corresponding to topics, 
by approximating document-word matrix as a product of document-factor and factor-word matrices.
Especially the NMF model has emerged as a popular alternative to probabilistic topic models 
\citinl{Arora2012, Choo2013, Greene2015, Brbic2016}, and evaluation experiments suggest
that its quality could be comparable to or better than the quality of the LDA model \citinl{ocallaghan2015}. 
As with the generative LDA model, there exist structural variations and extensions of the basic NMF model \citinl{Wang2013, Bai2018}.

In recent years neural topic models based on deep neural networks emerged as a popular approach \citinl{zhao2021topic}. 
Neural topic models have several appealing characteristics, 
including the automatization of the inference process and the ease of architectural extension,
the possibility of integration with other neural architectures, and scalability \citinl{zhao2021topic}.
This makes neural topic models better suited than the conventional topic 
models for tasks such as text generation, document summarization, and machine translation \citinl{zhao2021topic}.

\subsection{Topic Model Evaluation}

Topic models are practical since they are unsupervised and 
require no labeled data and minimal amount of text preprocessing. 
However, usefulness of topic models depends on the quality of the learned topics, 
which can vary and can be influenced by a multitude of factors.
Namely, deployment of a topic modeling solution involves choosing the model type, 
model hyperparameters, learning algorithm, and the preprocessing method.
In addition, once these choices are made, the process of model inference is stochastic, 
since the learning algorithms are initialized with random data and in many cases the learning process is based on random sampling.
Automatic evaluation of topic models can be used both to choose a better topic modeling approach
by narrowing down many available options, and to select model instances with high quality. 

A range of methods that evaluate various aspects of model and topic quality have been developed.
The earliest evaluation approach relies on measures of probabilistic fit that compute how well the learned model fits the data.
The most prominent measure of that type is perplexity of held-out data \citinl{Blei2003, Wallach2009}.
Perplexity was used in seminal topic modeling paper \citinl{Blei2003} and 
for many years remained a principal method for evaluation of newly proposed topic models.
Another probabilistic method, proposed by \citnoun{Mimno2011a}, measures the divergence 
between the learned model's latent variables and empirically estimated properties of these variables. 

An influential paper of \citnoun{Chang2009} demonstrated that lower perplexity of held out data 
does not neccesarily correlate with the interpretability of model topics.
These findings inspired an approach focused on directly quantifying topic interpretability
by calculating topic coherence \citinl{Newman2010}.
Measures of topic coherence compute a score that aims to approximate 
how interpretable a topic is in terms of its correspondence to a concept \citinl{Newman2010}, 
and are designed to achieve high correlation with human coherence assessments \citinl{Roeder2015}.
Coherence measures are commonly based on a score of mutual similarity between top topic-related words,
which can be defined using a variety of word representations and similarity measures 
\citinl{Newman2010, Mimno2011, Aletras2013, Lau2014, Rosner2014, Roeder2015, ocallaghan2015, Nikolenko2015, Nikolenko2016}.
Alternate approaches include clustering of word embeddings \citinl{Ramrakhiyani2017} 
and querying search engines with top topic words \citinl{Newman2010}.
In addition to topic coherence measures, alternate approaches to calculating topic quality have been proposed, 
based on calculating distances between topics and uninformative probability distributions \citinl{AlSumait2009},
and on aligning model topics with WordNet concepts \citinl{Musat2011, xu2020}.
A recent paper showed that the measures of topic coherence do not 
correlate well with the ability of humans to interpret and label topics \citinl{doogan2021},
and that the coherence measures are not a reliable guide for model selection \citinl{doogan2021}.
This experiment demonstrates the need for validation of automatic 
measures of model quality, which is rarely performed.   

The quality of topic models can also be assessed using human \hl{judgments} in a structured way. 
\citnoun{Chang2009} proposed a method for scoring semantic quality of topics using crowd-sourced answers to intrusion queries. 
Annotators were asked to choose an irrelevant word from a set of words describing a topic, 
or to choose a topic irrelevant to a document \citinl{Chang2009}.
\citnoun{ying2019inferring} extend the method of \citnoun{Chang2009} by proposing new intrusion tasks for topical quality,
as well as new tasks designed to measure correspondence between a topic and its conceptual label.

An approach to evaluation focused on model stability is motivated by 
inherent stochastic variability of learned model instances and 
the intuition that the consistency of learned topic is a desired property of a good model.
A common approach is to quantify stability as average mutual similarity of a number of model instances 
\citinl{DeWaal2008, Greene2014, Koltcov2014, Belford2018}. Model similarity has been calculated
using topic alignment based on bipartite matching \citinl{DeWaal2008, Greene2014, Koltcov2014, Belford2018} 
or directly comparing models using representations based on either words or documents \citinl{Belford2018}.
An approach to analyzing stability based on visualization of topic clusters of many model instances has been proposed in \citnoun{Chuang2015}.

Finally, if the information derived from a topic model is used as input for solving 
a downstream language processing task, a natural evaluation method is 
to quantify how much this information improves the performance on the task in question.
This approach is exemplified by applications of topic models for tasks 
such as information retrieval \citinl{Wei2006}, word sense disambiguation \citinl{boyd2007probabilistic}, 
sentiment analysis \citinl{titov2008joint}, and document classification \citinl{Stevens2012}.

\subsection{Topic Coverage}

The problem of topic coverage was first outlined in an article 
describing a framework for visual analysis of correspondence 
between expert-defined reference concepts and model topics \citnoun{Chuang2013}. 
The reference concepts were compiled by information visualization experts that relied on
domain knowledge and an indexed database of scientific articles. 
Matching of concepts and model topics is performed using a model that approximates
the probability that a human will judge a concept and a topic to be equivalent.
Several types of relations between concepts and model topics are defined. 
A concept is defined as resolved if it corresponds to a single topic, as fused if it is subsumed by a topic together with another concept, 
and as repeated in case it corresponds to multiple topics.
A concept is considered covered if it corresponds to at least one model topic, directly or as a part of a fused topic.
In a series of experiments, topic model types and hyperparameters are varied.
The resulting variations in relations between concepts and topics are presented by using the proposed visualizations tools. 
Finally, the alignment between concepts and topics is used to assess several measures of topic quality, including coherence measures.
We note that while our work is focused exclusively on topic coverage, the other types of relations between reference 
and model topics defined in \citinl{Chuang2013} are \hl{useful} tools for model analysis that merit further investigation.

Although there is no follow-up work to \citinl{Chuang2013} that focuses on topic coverage, there is work on related ideas.
In \citinl{chuang2014caca}, the authors analyze applications of topic models in social sciences, 
point to the problem of topic coverage, and argue that human-in-the-loop topic modelling 
might lead to models that best satisfy user needs.

A method of visual analysis of model stability proposed in \citinl{Chuang2015} is based on clustering
similar topics of many models and visualizing the relation between the models and the topic clusters.
Since the topic clusters can be viewed as reference topics, 
the visualizations in effect depict the random variations in coverage of a number of model instances. 

In \citnoun{Shi2019} authors propose a topic model analysis based on generating synthetic texts from a set of predefined synthetic topics.
From the perspective of topic coverage, such topics can be seen as reference topics, and be used 
for synthetic coverage experiments, possibly in conjunction with the readily deployable \cdc measure.
The use of synthetic topics could allow for large-scale analysis of numerous topic modelling scenarios 
without the need for manually crafting reference topics.
In \citinl{Shi2019}, the synthetic topics are not directly matched to model topics. 
Instead, the alignment between the two topic sets is computed indirectly, 
as the mutual information calculated on the level of words assigned to individual topics. 

Experiments in \citnoun{may2015topicid} evaluate several topic models on the tasks of topic identification and topic discovery.
Topic identification is defined in terms of the ability of the model-induced document-topic vectors 
to serve as features for classification and regression. 
Topic discovery is tested by measuring the alignment between the model-induced topics and gold-standard topic labels of documents.
This alignment is calculated as the similarity of the two partitions of documents, 
one induced by the model topics and the other induced by the gold-standard labels.
This approach is similar to the one in \citinl{Shi2019}, where the alignment between two topic sets is calculated indirectly, but at the word level.
It would be interesting to examine how these indirect measures relate to 
the coverage measures that directly match model topics to the reference topics represented in terms of word and document lists.

One approach to measuring topic quality is to align model topics to ontology concepts and 
define the quality score of a topic in terms of topic-concept relations \citinl{Musat2011, xu2020}. 
From the perspective of coverage, these techniques might prove useful for
the reverse task of measuring how the model topics cover concepts in large ontologies.
\citnoun{pandur2020} explore the similarities between model topics and the categories of the Web of Science taxonomy, 
and point out the problem of comparison between human- and model-generated taxonomies.
We believe that one way to approach this problem is from the perspective of coverage of taxonomy concepts.
The problem of coverage of abstract and broad concepts in both ontologies and taxonomies 
might prove interesting and challenging because of the need to conceptualize the 
relation between these concepts and model topics which tend to be more specific.

A big advantage of the coverage approach is its applicability 
for automatic analysis and validation of other measures of model quality.
Namely, the amount of work on methods for topic model evaluation is modest in comparison
to the amount of research on topic model architectures and applications, 
and the problem of semantic validation of topic models is far from solved. 
The automatic evaluation methods, spearheaded by popular coherence measures, 
are often used to compare a new topic model against a baseline model.
While they may be useful for providing a proof-of-concept for new model architectures, 
coherence measures are not reliable tools for guiding model selection in applications 
that rely on topic models for text analysis \citinl{ying2019inferring, doogan2021}.
The methods based on human inspection of topics \citinl{Chang2009, ying2019inferring} 
may provide more reliable assessments but they are time-consuming and rely on the availability of human annotators.

The coverage approach, unlike the measures of the abstract qualities of coherence and stability, 
is grounded in a set of interpretable reference topics representing a concrete application scenario of topic discovery.
Furhermore, the approach has a potential to lead to creation of many evaluation datasets, 
each consisting of a text corpus and a set of reference topics.
This would enable automatic testing of new topic models in varying topic discovery scenarios,
while the measures of model quality could be tested for their ability to select high-performing models.
Our research provides measures of coverage, datasets, and software tools that are a starting point for such analyses.

Experiments based on our methods confirm the previously detected unreliability of the coherence measures, 
and demonstrate the unrelatedness of topic model stability and coverage.
These findings underline the need for future work on improving and understanding the measures of model quality.

\section{Conclusions and Future Work} 

Topic models are a widely used tool for text exploration, often used for topic discovery on large text collections.
This paper explores an approach to topic model evaluation focused on measuring
to what extent topic models cover a set of reference topics -- representative set of topics of interest in a specific topic discovery scenario.

Our work revisits and extends the approach first outlined in \citnoun{Chuang2013},
by introducing new, reliable, and practical measures of coverage 
and performing a series of experiments on two different text domains, news and biological.
\hl{The measures we propose are the most important contribution of the paper 
since they make future coverage experiments more reliable and easier to perform.}
\hl{Our experiments} lead to findings about both topic models and other methods of topic model evaluation.
\hl{The} findings \hl{about topic models} include recommendations for the choice of models for topic discovery,
the experiments showing how the number of model topics influences coverage, 
and the demonstration that models' coverage depends on the semantic category of reference topics.
Experiments comparing topic coverage with topic coherence and model stability show that 
standard measures of coherence and stability fail to detect high-coverage models consistently and reliably.
These experiments underline the need for re-assessment and improvement of currently 
popular approaches to topic model evaluation.
We also show how the coverage measures can be successfully adapted to calculate model stability. 
\hl{Therefore, we demonstrate} that these measures are useful tools 
for matching models and topics\hl{, with} applications beyond the coverage-based evaluation.

\hl{The} most applicable contributions of our work are 
the \cdc measure of coverage and the recommendations for use of topic models in topic discovery.
The unsupervised \cdc measure is a new concept and a quickly deployable tool for model selection
that correlates very well with the coverage measure based on supervised topic matching.
The \cdc measure is based on the coverage-distance curve, which is in itself 
a \hl{useful} tool for graphical analysis and comparison of topic models.
For example, the \cdcurve{} can be used to assess and compare the levels of precision with which different models uncover the reference topics.
In addition, the \cdc measure has applications beyond coverage, since it can be used to \hl{assess} similarity of topic model instances.
Namely, the stability experiments show that the stability based on the \cdc measure
correlates almost perfectly with a standard stability measure. 

As for the recommendations for the applications of topic models for topic discovery, 
the results of the experiments indicate that the \nmfmodel model is a very good choice, 
having good performance on both text domains and outperforming probabilistic models in many cases.
The results of the experiments also support the use of models with a large number of topics.
Such \hl{models} have high coverage scores and are able to cover reference topics of all sizes.
On the other hand, the smaller models have poor coverage of small reference topics that can represent useful concepts.

The development of the coverage approach is still in the early stages 
and there exist many directions for future research.
One set of directions for future \hl{research} is related to the improvement of the \hl{measures} proposed in this paper.
The proposed supervised coverage measure relies on a time-consuming process of topic pair labeling.
We believe that active learning approaches \citinl{settles2009active} have the potential to greatly speed up this process.
The unsupervised \cdc measure performs well for model ranking and selection, 
but it could be further improved by making the computed coverage scores interpretable.

\hl{An important future research direction is the development of methods that facilitate} the construction of reference topics. 
\hl{Namely, reference topics are a key element of a coverage experiment, 
but their construction is technically challenging and time consuming.}
\hl{Therefore such methods would greatly} facilitate 
the application of coverage-based evaluation in new topic modeling scenarios.
\hl{In our view, a promising approach would be to focus on} graphical tools that would help the analyst to 
either select and modify automatically generated topics, or to create new topics based on expert knowledge.
Such \hl{graphical tools} could include metrics and visualizations for profiling of reference topics.
\hl{Tools of this kind could also facilitate the construction of incremental versions of 
a reference topic set. Such evolving collections of reference topics 
could be used in scenarios where texts and topics change over time.}

Each of the coverage experiments in this article outlines a potential direction for follow-up future work.
In general, \hl{similar} experiments in new topic modeling settings, based on other corpora and types of models, 
would lead to more robust findings and recommendations.
Specifically, we believe the computational social science could benefit from coverage experiments 
targeted at discovering topic models able to cover reference topics that correspond to 
concepts of interest in concrete scientific topic-discovery use cases. 
Construction of reference topic sets should not represent a significant overhead effort in these scenarios,
since the interpretation and analysis of a number of model topics is \hl{routinely} performed as part of model validation. 

Experiments with coverage-based evaluation of various types of topic models
may identify model architectures with consistently high performance for different corpora and reference topic sets.
We hypothesize that high coverage could be achieved by approaches that rely on 
pooling and combining of many model instances \citinl{suh2018, belford2020}, 
models that explicitly model topic diversity \citinl{Xie2015}, 
and models that iteratively learn new topics not uncovered by the previous runs \citinl{suh2018}. 
\hl{Alternative approaches to topic modeling, such as the one based on a combination of dimensionality reduction and 
soft clustering} \hla{\citinl{rashid2019}} \hl{, might also achieve good coverage results.} 

A promising \hl{future direction} is the application of the coverage methods 
\hl{to} large scale automatic analyses of the underresearched and rarely validated measures of model quality.
Such analyses could lead to better understanding and improvement of these measures.
For example, it would be quite useful to find or develop coherence measures 
that can well approximate the coverage of specific types of reference topics.
This could lead to coherence measures that are interpretable, 
and which \hl{could} be used to approximate coverage without the need for pre-constructed reference topics.

\hl{Topic models can be applied for text classification, either as extractors of topical features} 
\hla{\citinl{rashid2019},} \hl{or as stand-alone classifiers} \hla{\citinl{ramage2009}.}
\hl{One interesting future work direction is the investigation of the relationship between coverage and classification accuracy.}
\hl{In addition to revealing the nature of this relationship,
such experiments might lead to coverage-based recommendations for the use of classifiers based on topic models.}
\hl{A dataset for such an experiment should combine a classification dataset with reference topics, 
and we view the definition of classification-relevant reference topics as the main challenge.}
\hl{Similar experiments could be performed for applications of topic models to other language processing tasks, 
such as information retrieval} \hla{\citinl{Wei2006}}, \hl{word sense disambiguation} \hla{\citinl{boyd2007probabilistic}}, 
\hl{and sentiment analysis} \hla{\citinl{titov2008joint}.}

\hl{Topic models have also been applied to non-text data, most notably 
for the analysis of natural images and genetic data} \hla{\citinl{Blei2012}}.
\hl{In these applications topics are uninterpretable and correspond to distributions over genes or low-level visual patterns.}
\hl{The uninterpretability of non-text topics entails two important challenges -- 
the definition of sensible reference topics and the definition of topic matching.}
\hl{A wide-coverage set of topics recognized by a number of different models might be a good starting point, 
as might be the unsupervised} \hla{\cdc} \hl{measure that avoids the topic matching problem.}
\hl{We believe that the adaptation of the coverage approach to non-text domains represents 
an interesting direction for future work with the potential to generalize the approach and make it more robust.}

In this \hl{paper} we \hl{propose} a definition of the coverage problem motivated by the use case of topic discovery
-- a reference topic is considered covered if a closely matching model topic exist.
We proposed measures in line with this definition and experimented with two sets of  
reference topics within the reach of the standard topic models. 
\hl{However, the measures and the reference topics, two key aspects of the coverage problem,} can be viewed in \hl{a} more general light.
\hl{For example, in order to obtain more approximate coverage measures, the definition} 
of topic matching could be loosened to include approximate semantic similarity. 
\hl{Semantic variation among the reference topics could also be factored in the measures' design.
It could be quite useful to design measures that favor the models that cover a diverse set of 
reference topics' subcategories and offer a better overview of the semantic space.}

\hl{There are many ways to define potentially useful reference topics.}
\hl{For example, concepts of interest in social sciences, such as the news 
issues and frames, could be used to define useful sets of reference topics.}
\hl{Reference topics could also correspond to} concepts 
derived from a multitude of existing ontologies or taxonomies.
Another possibility is to use user-defined reference topics 
\hl{representing domain-relevant concepts}, as exemplified by \citnoun{Chuang2013}.
\hl{We note that defining of semantic reference topics is challenging, 
since it requires both a sensible definition of a set of concepts  
and a method of deciding whether an individual topic is in line with the definition.}
\hl{More broadly,  reference} topics need not even correspond to human concepts 
but could be synthetic, such as the topics in the experiment of \citnoun{Shi2019}.

\hl{Alternative approaches to the coverage problem represent a promising direction for future work. 
This work will have to deal with technical and conceptual challenges, such as}
the definition and construction of sensible reference topics, 
the semantics of matching between the \hl{model-generated topics} and the reference topics, 
and the efficient construction of practical measures of coverage.

We believe that future work on topic coverage can lead to a better understanding of
the semantics of machine generated topics and to improved evaluation methods 
with the potential to reasses the quality of existing models and guide the design of new ones.

\begin{appendices}
\section{Construction of Reference Topics}
\label{app:reftopics}

In this appendix to Section \ref{section:reftopics} 
we describe the details of the process of construction of the reference topics.
The reference topics are based on models' topics
inspected, interpreted, and filtered by human annotators.
\hl{More precisely, the reference topics corresponds to concepts discovered in two previous topic discovery experiments.
Each of the concepts is based on human inspection of either individual model topics or topic clusters.
Therefore, the reference topics are constructed from the model topics used in the previous experiments, 
and the methods of their construction reflect how the corresponding concepts are related to the model topics.}

\paragraph{News reference topics}
Reference topics of the news dataset were derived from the topics of LDA models
built and inspected in a study focused on topical analysis of political news texts \citinl{Korencic2015}.
Three LDA models with $50$ topics and two LDA models with $100$ topics were used.
Model topics were inspected by humans and interpreted as concepts, referred to as \emph{themes} in \citnoun{Korencic2015}. 
Themes were introduced as a conceptual tool for distributed annotation of model topics by several annotators, 
and a single theme was allowed to correspond to more than one model topic.
A shared list of themes was constructed in the process, with each theme described
by a label, a short description, and a list of model topics corresponding to the theme \citinl{Korencic2015}.
Model topics that do not correspond to any theme are thus uninterpretable topics.

Each of the reference topics corresponds to one of the $133$ themes from \citinl{Korencic2015}.
The topic's word and document vectors were derived, in two steps, from the model topics corresponding to the theme.
In the first step at most two corresponding model topics were selected at random, 
and in the second step the topics' data was improved by human effort.
The goal of this improvement was to ensure that the data of a reference topic describes the corresponding theme well. 
Namely, in the original annotation process a model topic containing a tolerable degree of noise 
was allowed to be labeled as corresponding to a theme \citinl{Korencic2015}. 

The improvement was \hl{performed} by two annotators who inspected model topics associated with each reference topic.
Upon inspection, they selected a subset of top topic words and documents that describe the reference topic well. 
\hl{Additionally}, each reference topic was labeled with a preference label denoting
weather the topic is better described by the words, documents, or equally well by both.
This was motivated by the observation that some reference topics were clearly best described by associated words, and some by associated documents.

Finally, topic-word and topic-document vectors of a reference topic were constructed from the annotators' data using the following procedure.
The reference topic's document vector is simply a binary indicator vector describing the documents associated with the topic.
The reference topic's word vector is constructed from the associated words' data
merged with the document-related data to reflect the preference of either word or document descriptors.
First the binary bag-of-words vector $vec_{words}$ describing the topic's words is constructed. 
The document-related data is represented by the vector $vec_{tfidf}$, the average of the documents' tf-idf vectors.
The final word vector of the reference topic is constructed as a weighted sum $w_w \cdot vec_{words} + d_w \cdot vec_{tfidf}$.
If the topic is best described by words, weights were set as $w_w=0.8$ and $d_w=0.2$.
Otherwise, if topic is best described by documents, the weights were $w_w=0.2$ and $d_w=0.8$,
and if there is no preference the weights were $w_w=0.5$ and $d_w=0.5$.

\paragraph{Biological reference topics}

Biological reference topics are based on the results of topic discovery performed 
with the goal of finding topics corresponding to phenotypes -- characteristics of organisms \citinl{Brbic2016}.
The original topic discovery process was a part of a set of machine learning methods 
developed with the purpose of large scale annotation of organisms with corresponding phenotypes \citinl{Brbic2016}.

The original topic discovery was performed by human inspection of clusters of topics 
of NMF models built from biological texts describing microorganisms. 
One NMF model with $50$ topics and one NMF model with $100$ topics were built for
each of the five subcorpora corresponding to texts of five text sources described in Section \ref{section:corpora}.
Then the topics of the NMF models with same number of topics were clustered, 
using as the measure of similarity the Pearson correlation between sets of top topic words.
The clusters were then filtered by retaining only the clusters containing topics from at least three out of five subcorpora, 
guided by the requirement that a phenotype should be consistently uncoverable across text sources. 
In order to increase the coverage of phenotypes, additional clusters were generated
using the described procedure and new topic models were built with different random seeds.
In total, five models with $50$ topics and four models with $100$ topics were built for each of the text sources.

The obtained clusters were represented by averaging the topic-word vectors of the cluster's topics
and selecting the $20$ top-weighted words from the resulting vector \citinl{Brbic2016}.
Inspection and interpretation of these clusters was performed by a biologist who selected 
the high quality clusters, characterized by consistent and relevant words and corresponding to phenoptype concepts. 
This process resulted in a total of $112$ topic clusters corresponding to phenotypes.

The reference topics of the biological dataset are derived from the described topic clusters, 
and their topic-word and topic-document vectors are constructed in the following way. 
The topic-word vector of a reference topic is a binary bag-of-words vector of the corresponding cluster's words.
The topic-document vector of a reference topic is constructed by averaging the topic-document 
vectors of the corresponding cluster's topics.
These topic-document vectors are extracted from the NMF models obtained by the original topic discovery study.

\section{Construction of Topic Models}
\label{app:modelbuild}

Here we append Section \ref{section:covmodels} with details of topic models' construction.

For the construction of the probabilistic models (\ldamodel, \aldamodel and \pypmodel), 
we rely on the implementation of inference algorithms
provided as part of the HCA package \citinl{Buntine2014}. 
This software implements the optimized variant of Gibbs sampling named table indicator sampling \citinl{Chen2011sampling},
combined with adaptive rejection sampling \citinl{Gilks1992} for hyperparameter learning.
Following the standard procedure of \citnoun{Griffiths2004}, the hyperparameters of the \ldamodel model
defining the priors of the topic-document and topic-word distributions are set to $\alpha=50/T$ and $\beta=0.01$.
For the \aldamodel model, the $\beta$ hyperparameter is also set to $0.01$. 
Initial values of the Gamma distribution parameters defining the \aldamodel's prior document-topic distribution 
are set to $a=0.5$ and $b=10$, for each of the topics. 
For the \pypmodel model, the initial values of the concentration and discount parameters 
of the Pitman-Yor process are set to $c=10$ and $d=0.5$.
A large number of Gibbs sampling cycles is \hl{performed} since
in our case the goal of the learning process is the quality of learned models, not the speed of learning.
After $50$ warmup cycles of Gibbs sampling, another $800$ cycles are run in case of the \ldamodel and \aldamodel models, 
while in case of the \pypmodel model with more parameters, another $1500$ cycles are run. 

To construct the \nmfmodel model instances, we use the method described in \citnoun{Greene2015}.
Text documents are represented as a matrix of tf-idf document-word weights,
and the matrix factorization is performed using the projected gradient method \citinl{Lin2007}
initialized with the results of the non-negative SVD decomposition \citinl{Boutsidis2008}.
We use the implementation of the described method available 
as part of the scikit-learn framework \citinl{Buitinck2013}. 

\section{Balancing the Dataset of Topic Pairs}
\label{app:pairbalance} 

Here we append Section \ref{sect:topicpairdataset} with details of the problem of imbalance of topic pairs and its solution.

When a subset of topic pairs is randomly sampled from a set of all possible pairs containing random model topics,
a large majority of pairs will contain non-matching topics. 
Namely, in an ideal scenario with two models each having $T$ topics and where all the topics match one of $T$ distinct concepts, 
the probability of match of two randomly selected topics equals $1/T$.
In a realistic scenario with a large number of concepts and potentially noisy topics,
it is reasonable to expect that the probability of match of two topics will be below $1/T$, 
as was confirmed by an inspection of a sample of pairs.

This means that the topic matching problem falls in the domain of imbalanced learning \citinl{Branco2016, Krawczyk2016} --
a setting in which only a small fraction of positive learning examples is expected in the learning data.
This hinders learning of good models since examples that define the structure of the positive class are scarce.
Many approaches to alleviate and solve this problems were developed \citinl{Branco2016}, including active learning and resampling methods.

However, in case of the problem of topic matching there exists a simple solution --
using a measure of topic distance to sample a more balanced dataset.
The intuition behind the approach is that the distance between two topics is inversely correlated with the probability of their match.
Therefore, if pairs of mutually close topics are sampled with the same probability as the pairs of distant topics, 
the final sample is expected to contain more matching topics and thus provide a better dataset for model learning.

We use the cosine distance of topic-word vectors to measure topic distance, 
since this measure is able to approximate the human intuition of topic similarity reasonably well \citinl{Chuang2013}.
To create the balanced sample, the dataset of all topic pairs is partitioned into subsets corresponding to distance subintervals. 
Specifically, cosine distance between positive topic-word vectors ranges between $0$ and $1$,
and we partition the $[0,1]$ interval into $10$ subintervals of equal width.
Each subset contains the pairs of topics whose mutual distance falls within the corresponding interval's boundaries.
The final balanced sample of topic pairs is created by sampling the same number of pairs from each of the subsets.
Inspection of a validation sample showed that it contains $36\%$ 
of pairs with matching topics, as opposed to less than $1\%$ of matches expected from a fully random sample.

\section{Details of Topic Pairs Annotation}
\label{app:pairannot} 

Here we append Section \ref{sect:pairlabeling} by describing in detail the process of annotation of topic pairs. 
For the ease of reference we first repeat, in a compact form, the definition of a topic match,
and the definition of the labels used to annotate topic pairs.

A topic match is defined as conceptual equality of topics --
two topics are considered equal if they are interpretable as the same concepts, 
where the interpretation of a topic as a concept is as specific as possible.
On the semantic level, we define topic equality as matching of
concepts obtained by interpreting topics as specifically as possible.
Matching of concepts is defined as equality or near equality of concepts, allowing small variations and similar aspects of a same concept.
Stochastic differences are accounted for by labeling topics as equal but with presence of noise.
This is the case when one or both topics contain a noticeable amount of noise but the topics are still
interpretable and the equality of interpreted concepts exists as \hl{previously} defined.

A pair of topics is labeled with $1$ in case of topic equality, i.e., when concepts match without noise. 
A pair is labeled with $0.5$ in case of a match in the presence
of noise or small semantic variation, and with $0$ when the concepts do not match.

News topic pairs were annotated by the authors that performed topic discovery and analysis \citinl{Korencic2015}
on the corpus from which the news reference topics were derived, and by students of English studies acquainted with the topics of US politics.
Pairs of biological topics were labeled by a biological \hl{scientist} and students of senior years of biology.

Precise labeling instructions were formed, containing the previous definition of a topic match, 
examples of topic pairs, and clarifications of the labeling process.
Annotators proceeded to annotate the previously described dataset of topic pairs containing both model and reference topics. 
Each topic was represented as a list of $15$ top-ranked topic words and $15$ top-ranked topic documents. 
Documents were represented as informative summaries -- titles of news articles and initial fragments of original text in case of biological texts. 
The annotators also had access to full text of the documents.

\newcommand{\alphao}{$\alpha_{o}$\xspace}
\newcommand{\alphan}{$\alpha_{n}$\xspace}

The process of annotation was performed according to the instructions from \citinl{Krippendorff2012}.
For each dataset, in each round of annotation all the topic pairs were annotated by three annotators.
Annotation quality was \hl{assessed} using Krippendorff's $\alpha$ coefficient that 
measures mutual agreement of the annotators corrected for the possibility of random agreement \citinl{Krippendorff2012}.
Two versions of $\alpha$ coefficient were used -- nominal $\alpha$, which measures strict equality of annotations, 
and ordinal $\alpha$ based on the distance of annotations on the ordinal scale. 
The two versions are labeled as \alphan ande \alphao, respectively.

At the beginning of the annotation process, a small pilot set of 15 topic pairs 
was annotated by all the annotators in order to clarify the instructions.
In the next step a calibration set of $50$ topic pairs was annotated and the application of annotation
instructions was discussed for topic pairs with large disagreement.
The $\alpha$ coefficients of the calibration round were calculated, yielding \alphan of $0.568$ and \alphao of $0.831$ for news topics, 
and \alphan of $0.576$ and \alphao of $0.712$ for biological topics.
Next a test set of $50$ topic pairs was annotated, yielding \alphan of $0.663$ and \alphao of $0.862$ for news topics, 
and \alphan of $0.599$ and \alphao of $0.782$ for biological topics.
Improvements of the agreement coefficients were interpreted as a consequence of 
clarification of both the annotation instructions and the method of their application.
Lower annotator agreement for pairs of biological topics is likely a consequence of
the fact that biological topics, as opposed to news topics, correspond to more complex and 
abstract concepts that are harder to interpret. 

After the first two annotation rounds the agreement coefficients were deemed sufficiently high for both datasets.
This decision was \hl{additionally} supported by the feedback from the annotators who 
\hl{assessed} both the definitions of topic matching and the process of annotation as reasonable and comprehensible.
The annotation process was continued and for each dataset another $250$ topic pairs were annotated. 
These pairs were merged with the $50$ pairs form the test set to produce 
the final set of $300$ topic pairs, each pair annotated by three annotators.

For the final sets containing all the annotated pairs, the calculation 
of $\alpha$ agreement coefficients yielded \alphan of $0.689$ and \alphao of $0.865$ for news topics, 
and \alphan of $0.648$ and \alphao of $0.797$ for biological topics.

\section{Construction of the Supervised Topic Matcher}
\label{app:supmodels}  

\begin{table*}
\caption{Supervised models used for classification of topic pairs and models' hyperparameters that are optimized in the process of model selection.}
\centering
{\small
\begin{tabular}{llp{21em}}
\toprule
Model & Hyperparameter & Hyperparameter values \\
\midrule
Logistic regression & regularization constant & $0.001, 0.01, 0.1, 1.0, 10, 100, 1000$ \\
& regularization norm & \loned, \ltwod \\
\midrule
Multilayer perceptron & hidden layer width & $3, 5, 10$ \\
& regularization constant &  $0.00001, 0.0001, 0.001, 0.01, 0.1$ \\
\midrule
Random forest & number of trees &  $10, 20, 50, 100$ \\ 
& number of features & $2$, $50\%$, all features \\
& maximum tree depth & $2$, $3$, unlimited \\  
\midrule
Support vector machine & regularization constant & $0.001, 0.01, 0.1, 1.0, 10, 100, 1000$\\
& radial basis function \rbfgamma & $0.001, 0.01, 0.1, 1.0, 10, 100, 1000$, \autogamma \\  
\bottomrule
\end{tabular}}
\label{table:supcovmodels}
\end{table*}

Here we append Section \ref{sect:supmodelconstr} with the details of the methods
of feature construction and model construction. 
The goal of these methods is the construction of a binary classifier 
that predicts weather a pair of topics matches or not.

Four standard classifiers are considered: logistic regression \citinl{Murphy2012}, support vector machine \citinl{Cortes1995}
with radial basis function kernel, random forest \citinl{Breiman2001}, and multilayer perceptron \citinl{Murphy2012}.
We use the implementations of the models available as part of 
the scikit-learn framework \citinl{Buitinck2013}. 
Classification models and the corresponding hyperparameters that we optimize are summarized in Table \ref{table:supcovmodels}.
Other hyperparameters are set to sensible default values defined by the scikit-learn framework \citinl{Buitinck2013}.

In order to perform supervised classification of topic pairs, each pair is represented as a vector of features.
These features should contain information enabling a good approximation of semantic matching.
Preliminary experiments with features constructed by concatenating topic-word and topic-document vectors
of the topics in a pair resulted in a relatively low classification performance, yielding \fone scores between $0.4$ and $0.6$.
A plausible explanation for this result is the so called curse of dimensionality \citinl{Tan2005} -- 
degradation of classification accuracy caused by high dimensionality of feature vectors (in our case, tens of thousands) 
and a small number of learning examples (in our case, a few hundreds).

A possible solution for this problem is feature extraction \citinl{Tan2005} -- transformation of high-dimensional representations
into small feature vectors containing useful information. 
Previous experiments with topic models show that distance measures applied to topic-word vectors 
can be used to approximate semantic similarity of topics \citinl{Zhao2011, Chuang2013, Roberts2015, Chuang2015}.
Preliminary experiments with features based on various distance measures 
applied to topic-word and topic-document vectors showed promising results, yielding \fone scores between $0.7$ and $0.8$.
Therefore, we opt for this approach to feature extraction.

We base the features representing a pair of topics on the following four distance measures:
cosine distance, Hellinger distance \citinl{Jebara2003}, \loned distance, and \ltwod distance.
These four measures represent four distinct measure types: angular distance, distance between probability distributions, 
and two standard measures of coordinate distance \loned and \ltwod.

Before the application of a distance measure, topic-word and topic-document vectors are normalized to probability distributions.
This is \hl{necessary} in order for the Hellinger distance to be applicable and, 
in the case of \loned and \ltwod, to insure the insensitivity of features to the type of topic models.
Namely, probabilistic topic models produce topic-word and document-topic vectors that contain 
small values corresponding to probabilities, while the \nmfmodel topic model produce vectors of unbounded and potentially large positive values. 
Therefore, unnormalized features would result in distance variation that reflects the difference in topics' types.

The final feature representation of a topic pair is constructed 
by applying the previous four distance measures to both the pair of normalized topic-word vectors
and to the pair of normalized topic-document vectors. 
This way each topic pair is represented with eight features, 
four based on topic-related words and four based on topic-related documents. 

For each of the four classification models, we use the entire dataset of $300$ labeled topic pairs
to assess the performance of the model variant with optimized hyperparameters.
The \hl{assessment} is done using the procedure of nested five fold crossvalidation 
and the \fone measure is used to measure the classification performance of the models.

\begin{figure*}[h!]
\centering
\includegraphics[width=2\columnwidth]{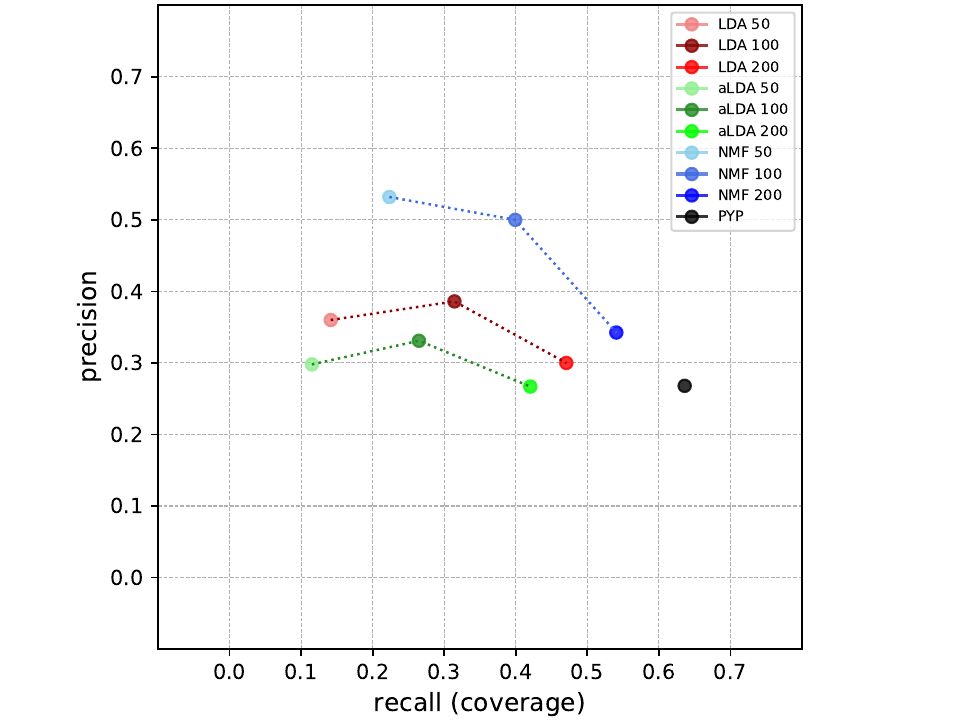}
\caption{\footnotesize \hl{Models' precision (the fraction of model topics in the reference set) 
and recall (the fraction of reference topics covered), on the news dataset.}} 
\label{fig:pruspol}
\end{figure*}

\begin{figure*}[h!]
\centering
\includegraphics[width=2\columnwidth]{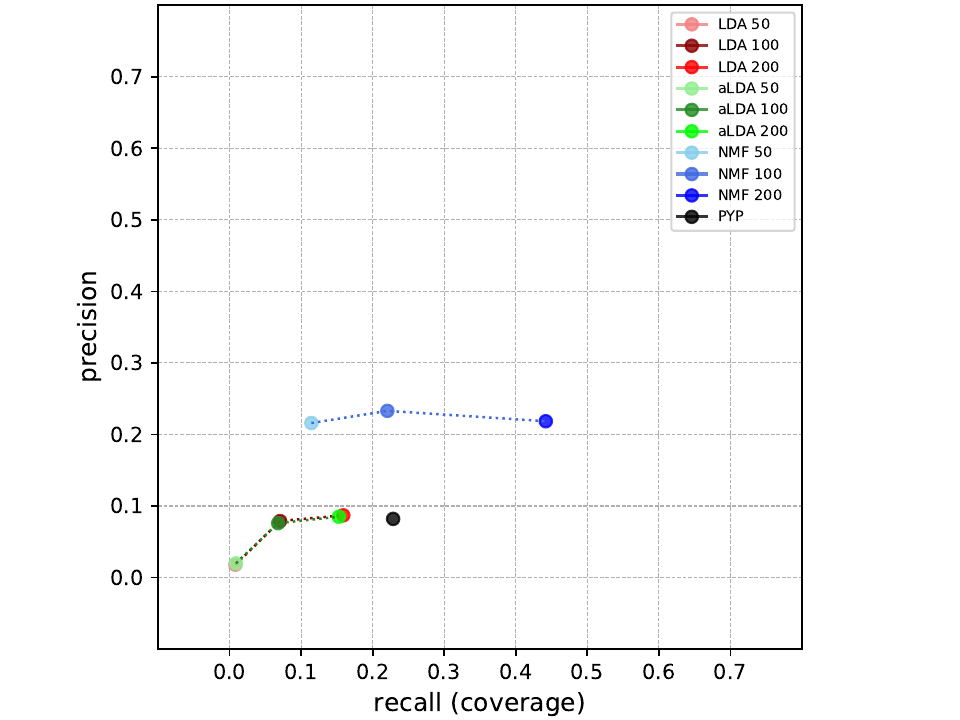}
\caption{\footnotesize \hl{Models' precision (the fraction of model topics in the reference set) 
and recall (the fraction of reference topics covered), on the biological dataset.}} 
\label{fig:prpheno}
\end{figure*}

When performing the standard non nested crossvalidation with $K$ folds, model performance obtained for 
each combination of hyperparameters is calculated by learning the model on $K-1$ folds 
(distinct subsets of the learning data), and calculating the performance on the remaining fold. 
The final quality score is obtained as the average over all $K$ folds.
When performing nested crossvalidation, for each ``outer'' subset of $K-1$ folds, 
full hyperparameter optimization is performed using non nested crossvalidation 
which partitions the subset into $K$ ``inner'' folds.
In other words, nested crossvalidation uses regular crossvalidation to assess
the entire process of hyperparameter optimization, not just to assess one combination of hyperparameters.
Athough computationally more expensive, nested crossvalidation gives better 
assessment of the quality of a model obtained by hyperparameter optimization \citinl{Cawley2010}.
We generate crossvalidation folds using stratified sampling in order to preserve, 
for each fold, the ratio of class labels that is representative for the entire dataset.

The described methods leads to optimized models that achieve an \fone 
score of approximately $0.8$, with variations in performance that depend on the model and the dataset.
The logistic regression model achieves best \fone scores, and the classifiers' performance 
is close to the mutual agreement of human annotators.

\section{Supplementary Coverage-related Experiments}
\label{app:covapp} 

\hl{Here we supplement the Section} \hla{\ref{sect:modelcoverage}} \hl{with an 
analysis of models' precision and recall, and with an empirical analysis 
of the running time of the coverage measures.}

\subsection{Relationship Between Model Precision and Recall}

\hl{
We define the precision and recall of a topic model in terms of the 
relevant topics (topics matching the reference topics) retrieved by the model.
A topic model's recall -- fraction of the reference topics retrieved by the model -- is equal to the model's coverage.
Model precision is the fraction of the relevant model topics -- model topics that match the reference topics.
If more than one model topic matches the same reference topic, only one model topic is counted as relevant.
However, such redundancy does not occur in our experiments -- for each topic model instance 
a retrieved reference topic is always matched by a single model topic.
This might seem counterintuitive since it is, at least in our experience, not unusual that a model 
contains mutually similar topics.
The explanation is that our supervised matcher, described in Section} \hla{\ref{section:supmeasures},}
\hl{is built to match only highly similar topics.}

\hl{In this experiment we analyze the precision and recall of topic models 
analyzed in the coverage experiments of Section} \hla{\ref{sect:modelcoverage}.}
\hl{The details of these models of various types and sizes are described in Section} \hla{\ref{section:covmodels}.}
\hl{Similarly as in the coverage experiments, for each combination of a model type and a number of topics 
precision and recall scores of the $10$ distinct model instances are averaged.}

\hl{The results are shown in Figure} \hla{\ref{fig:pruspol}} \hl{and Figure} \hla{\ref{fig:prpheno}.}
\hl{Relation between precision and recall depends on both the dataset and the model type.
One might expect that as the recall (coverage) rises with the increase of the model size, 
the precision (the proportion of the relevant model topics) will decline.
However, this tradeoff occurs only in some cases, and it does not entail a large loss of precision.
For most topic model types, the increase in the number of topics 
is related to an insignificant decrease or even to a small increase in precision.
On the biological dataset, in most cases the precision remains stable as the recall increases.}
\hl{The tradeoff is most noticeable in case of the} \hla{\nmfmodel} \hl{model on the news dataset. 
However, even in this case there is no drastic loss of precision -- the}
\hla{\nmfmodel} \hl{model with $200$ topics more than doubles the recall of the} \hla{\nmfmodel} 
\hl{model with $50$ topics, while the corresponding loss in precision is only $35\%$.}

\hl{The results of this experiment support the use of models with a larger number of topics,
which is in line with the experiments in Sections} \hla{\ref{sect:modelcoverage}} \hl{and} \hla{\ref{sect:covbysize}.}
\hl{Namely, larger models offer a significant increase in coverage (recall),
which rarely comes at a price of a noticeable loss of precision.
The} \hla{\nmfmodel} \hl{model is better then the probabilistic models in terms of precision as well as in terms of recall.
In other words, the} \hla{\nmfmodel} \hl{instances will expectedly contain more relevant topics, 
which should lead to quicker topic discovery.}
\hl{This is in line with the previous recommendations for the use of the} \hla{\nmfmodel} 
\hl{model from Section} \hla{\ref{sect:modelcoverage}.}

\hl{While the larger models do not suffer a large loss of precision, 
the absolute number of their topics outside the reference set is higher then in the case of smaller models.
Therefore an analyst might perceive larger models as less useful. 
This observation is in line with the recommendation form Section} \hla{\ref{sect:covbysize}}
\hl{that tools that speed up the process of topic inspection should be used in conjunction with large models.}

\subsection{Running Time of the Coverage Measures}

\begin{table*}[h]
\caption{\footnotesize \hl{Running time (in seconds) of the supervised} \hla{\sups} \hl{measure and the unsupervised} 
\hla{\cdc} \hl{measure -- time required for the processing of the entire set of $100$ topic models, 
average time per model topic, and average time per model. The experiment was performed using a 2.4 GHz Intel i7 processor.}} 
\label{table:measuretime}
\centering
\normalsize
\begin{tabular}{lrrrrrr}
\toprule
             & \multicolumn{3}{c}{News dataset} & \multicolumn{3}{c}{Biological dataset} \\            
\cmidrule(lr){2-4} \cmidrule(lr){5-7}
             & all models & topic avg. & model avg. & all models & topic avg. & model avg.  \\
\midrule         
\sups & 4707.43 & 0.349 & 47.07 & 1149.31 & 0.0851 & 11,49 \\
\cdc & 39.10 & 0.0029 & 0.391 & 8.64 & 0.00064 & 0.0864 \\
\bottomrule
\end{tabular}
\end{table*}

\hl{Time complexity of the coverage measures influences the scalability of the coverage experiments.
Asymptotic complexity of the proposed coverage measures is analyzed in Section} \hla{\ref{section:measures}.}
\hl{In this section we perform an empirical analysis of the measures' running time. 
For each of the datasets, we timed the calculation of the coverage measures 
on the set of $100$ topic models described in Section} \hla{\ref{section:covmodels}.}

\hl{In terms of the number of reference topics $R$, the number of model topics $T$, 
the vocabulary size $V$, and the corpus size $D$, the complexity of the supervised}
\hla{\sups} \hl{measure is $\mathcal{O}(R T (V+D))$, while the complexity of the unsupervised}
\hla{\cdc} \hl{measure is $\mathcal{O}(RTV)$.
The asymptotic complexities might wrongly suggest similar running times,
especially since for both datasets the vocabulary size is very close to the corpus size.
However, the results, displayed in Table} \hla{\ref{table:measuretime},} \hl{show that in practice
the unsupervised} \hla{\cdc} \hl{measure is two orders of magnitude faster then the supervised} \hla{\sups} \hl{measure.}
\hl{This is caused by the fact that the time required to process a pair of topics differs greatly between the measures.}
\hl{For the} \hla{\sups} \hl{measure, the processing of a topic pair requires the calculation of 
eight distance-based features and the computation of the supervised model's output.
In contrast, in case of the} \hla{\cdc} \hl{measure the only operation required is the calculation of cosine distance.
Additionally, for the} \hla{\cdc} \hl{measure, the distances between all topic pairs are pre-computed using matrix-level computation.
This is more efficient than calculating vector distances for each pair of topics.}

\hl{The results show that the} \hla{\cdc} \hl{measure is time-efficient and well suited for large experiments.
However, the} \hla{\sups} \hl{measure could be optimized by using less features or a more efficient supervised model, 
assuming that this would not cause a degradation in accuracy.
Another possible optimization is to pre-compute the distance-based features using matrix-level operations.
Additionally, the computation of both measures could be parallelized by 
distributing the topic models across the available processor cores.}

\end{appendices}

\section*{Acknowledgment}

We would like to thank Maria Brbi\'c for the patient help with the 
technical details related to the data from \citinl{Brbic2016}.
We would also like to thank Mladen Karan for the help with the server used to conduct part of the experiments.

\bibliographystyle{IEEEtran}
\bibliography{bibliography}

\begin{IEEEbiography}[{\includegraphics[width=1in,height=1.25in,clip,keepaspectratio]{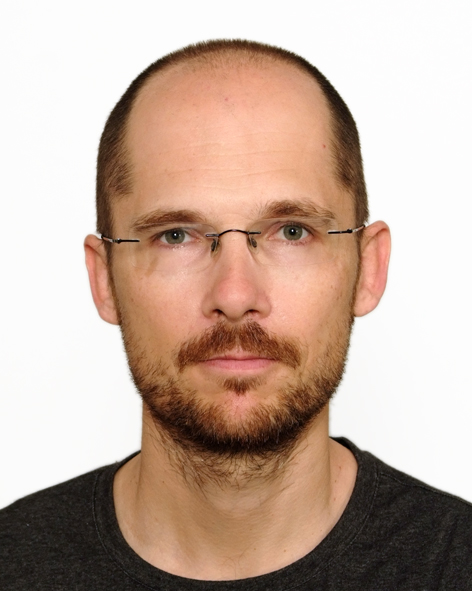}}]{Damir Koren\v{c}i\'{c}} 
was born in Zagreb, Croatia, in 1983. He received the B.S./M.S. degree in mathematics in 2008 
and the Ph.D. degree in computer science in 2019, both from the University of Zagreb.

From 2010 to 2016 he was a Research Assistant at the Ru{\dj}er Bo\v{s}kovi\'{c} Institute, 
and from 2018 to 2019 he was a Research Assistant at the Faculty of Electrical Engineering and Computing, University of Zagreb.
Since 2019 he has been a Postdoctoral Researcher at the Department of Electronics, 
Ru{\dj}er Bo\v{s}kovi\'{c} Institute in Zagreb, Croatia.
He also worked as a Software Engineer, and as a Teacher at both the high school and the university level.
His research interests include text mining, topic modeling, 
text analysis for computational social science, and data compression.

\end{IEEEbiography} 

\begin{IEEEbiography}[{\includegraphics[width=1in,height=1.25in,clip,keepaspectratio]{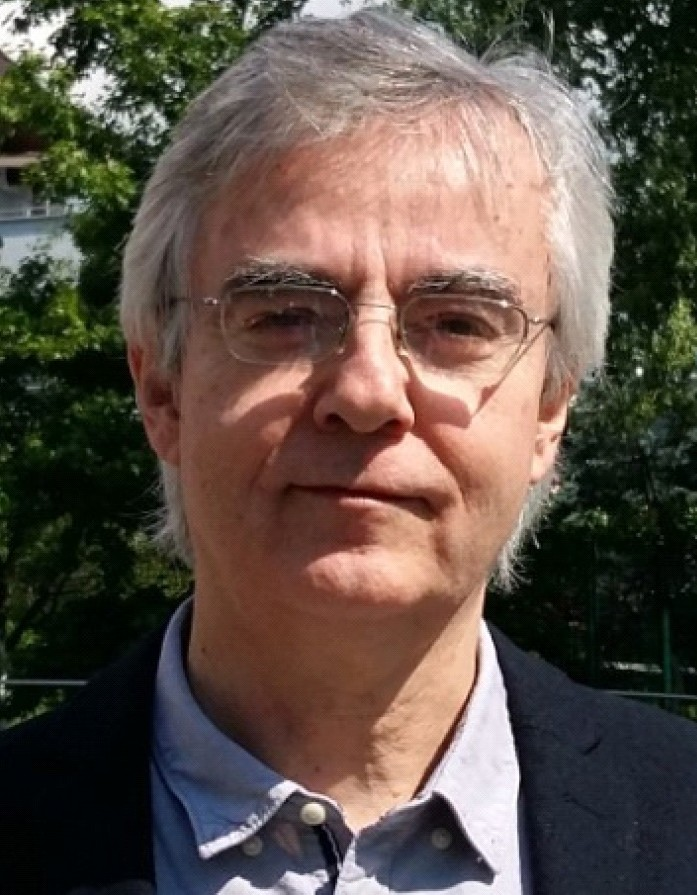}}]{Strahil Ristov} 
was born in Zagreb, Croatia, in 1959. He received his B.S. degree in electrical engineering and M.S. and Ph.D. degrees 
in computer science from the University of Zagreb in 1997. 

Since 1990, he has been a Researcher at the Department of Electronics, Ru{\dj}er Bo\v{s}kovi\'{c} 
Institute in Zagreb and currently holds the position of Senior Associate Scientist. 
He is the head of the Laboratory for Information and Signal Processing. 
He is the author or coauthor of 30 journal or conference papers. 
His research interests include string algorithms, data compression, and algorithms in bioinformatics and population genetics.
\end{IEEEbiography} 

\begin{IEEEbiography}[{\includegraphics[width=1in,height=1.25in,clip,keepaspectratio]{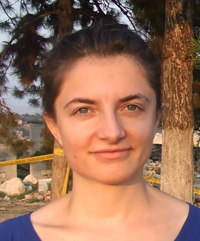}}]{Jelena Repar} 
was born in Croatia, in 1982. She received the B.S./M.S. degree in molecular biology (2006.) 
and the Ph.D. degree (2012) in biology from the University of Zagreb.

From 2006 to 2012 she worked as a Research Assistant, and from 2012 to 2014 as a Senior Research Assistant at the 
Division of Molecular Biology, Ru{\dj}er Bo\v{s}kovi\'{c} Institute, Zagreb, Croatia. 
From 2014 to 2017 she worked as a Postdoctoral Researcher at MRC London Institute of Medical Sciences, Imperial College London, UK. 
Since 2018 she has been an Associate Scientist at the Division of Molecular Biology, Ru{\dj}er Bo\v{s}kovi\'{c} Institute, Zagreb, Croatia. 
Her research interests include microbes, DNA repair, genomics and computational biology.
\end{IEEEbiography} 

\begin{IEEEbiography}[{\includegraphics[width=1in,height=1.25in,clip,keepaspectratio]{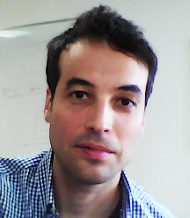}}]{Jan \v{S}najder} 
was born in Zagreb, Croatia, in 1977. He received his B.S. degree in computing in 2001 and M.S. and Ph.D. degrees in computer science from the University of Zagreb, in 2006 and 2010, respectively.

Since 2001 he has been a Researcher at the Department of Electronics, Microelectronics, Computer and Intelligent Systems at the Faculty of Electrical Engineering and Computing, University of Zagreb, where he currently holds the position of an Associate Professor. He is the author or coauthor of over 100 journal or conference papers. His research interests include natural language processing, with a focus on information extraction and text analysis for computational social science.
\end{IEEEbiography}

\EOD

\end{document}